\documentclass[twocolumn,prl,sort&compress,floatfix]{revtex4-1}
\usepackage{amssymb}
\usepackage{amsmath}
\usepackage{mathrsfs}
\usepackage{bm}
\usepackage[dvipsnames]{xcolor}
\usepackage[sort&compress]{natbib}

\usepackage{graphicx}

\usepackage{color}
\definecolor{red}{rgb}{1,0,0}
\definecolor{blue}{rgb}{0,0,1}

\renewcommand{\figurename}{\textbf{Figure}}

\begin{document}
\title{Dynamical Scaling and Phase Coexistence in Topologically-Constrained DNA Melting}

\author{Y. A. G. Fosado$^{1,*}$, D. Michieletto$^{1,*}$ and D. Marenduzzo$^1$}
\affiliation{$^1$ SUPA, School of Physics and Astronomy, University of 
	Edinburgh, Peter Guthrie Tait Road, Edinburgh, EH9 3FD, UK. $^*$ Joint first author \\ $^\dagger$For correspondence: D. Michieletto (davide.michieletto@ed.ac.uk), D. Marenduzzo (dmarendu@ph.ed.ac.uk)}

\begin{abstract}
	\textbf{There is a long-standing experimental observation that the melting of topologically constrained DNA, such as circular-closed plasmids, is less abrupt 
	than that of linear molecules. This finding points to an intriguing role of topology in the physics of DNA denaturation, which is however poorly understood.
	Here, we shed light on this issue by combining large-scale Brownian Dynamics simulations with an analytically solvable phenomenological Landau mean field theory. We find that the competition between melting and supercoiling leads to phase coexistence of denatured and intact 
	phases at the single molecule level. This coexistence occurs in a wide temperature range, thereby accounting for the broadening of the transition.  
	Finally, our simulations show an intriguing topology-dependent scaling law governing the growth of denaturation bubbles in supercoiled plasmids, which can be understood within the proposed mean field theory.
	 }
\end{abstract}

\maketitle

One of the most fascinating aspects of DNA is that its biological function is intimately linked to its local topology~\cite{Bates2005}. For instance, DNA looping~\cite{Alberts2014,Rippe1995} and supercoiling~\cite{Liu1987,Bates2005,Brackley2016scoil} are well-known regulators of gene expression, and a variety of proteins, such as Polymerases, Gyrases and Topoisomerases, can affect genomic function by acting on DNA topology~\cite{Alberts2014,Bates2005}. 


Fundamental biological processes such as DNA transcription and replication are associated with local opening of the double helix, a phenomenon that can be triggered  
\emph{in vitro} by varying temperature, pH or salt concentration~\cite{Wartell1985}.
The melting transition of DNA from one double-stranded (ds) helix to two single-stranded (ss) coils has been intensively studied in the past by means of buoyant densities experiments~\cite{Vinograd1968}, hyperchromicity spectra~\cite{Jensen1976}, AFM measurements~\cite{Jeon2010}, single-molecule experiments~\cite{Vlijm2015} and 
fluorescence microscopy~\cite{Altan-Bonnet2003}.

In particular, experiments~\cite{Vinograd1968,Wartell1985} and theories~\cite{Poland1966a,Kafri2000} have shown that the ``helix-coil'' transition in linear or nicked DNA molecules, which do not conserve the total linking number between the two strands, is abrupt and bears the signature of a first-order-like transition. On the other hand, the same transition is much smoother for DNA segments whose linking number is topologically preserved, such as circular, covalently-closed ones~\cite{Vinograd1968,Gagua1981}.

Understanding the physical principles underlying DNA melting in topologically constrained (tc) DNA is important since this is the relevant scenario {\it in vivo}. For instance,  bacterial DNA is circular, while in eukaryotes DNA wraps around histones~\cite{Alberts2014}, and specialised proteins are able to inhibit the diffusion of torsional stress~\cite{Naughton2013}. 
 
Intriguingly, and in stark contrast with the behaviour of linear, or topologically free (tf), DNA, the width of the melting transition of tcDNA is relatively insensitive to the precise nucleotide sequence~\cite{Gagua1981} thereby suggesting that a universal physical mechanism, rather than biochemical details, may underlie the aforementioned broadening.
While biophysical theories of tcDNA melting do exist, they do not reach a consensus as to whether the transition should weaken or disappear altogether~\cite{Rudnick2002,Kabakcioglu2009a}; it is also unclear if stable coexistence of ss (coiled) and ds (helical) phases may in general be possible regardless of sequence specificity~\cite{Benham1979,Sen1988,Bauer1993}. 

Here we address this topic by employing a combination of complementary methods. First, we perform large-scale coarse-grained Brownian Dynamics (BD) simulations of 1000 base-pairs (bp) long topologically free and constrained double-stranded DNA molecules undergoing melting~\cite{Fosado2016}. These unprecedentedly large-scale simulations at single bp resolution allow us to measure both kinetic and equilibrium observables of the melting transition. Second, we propose and study a phenomenological Landau mean field theory which couples a critical ``denaturation'' field ($\phi$) with a non-critical ``supercoiling'' one ($\sigma$). This approach captures the interplay between local DNA melting and topological constraints, and predicts the emergence of phase coexistence within a wide temperature range, in line with our simulations, and accounting for the experimental broadening of the melting transition in tcDNA. We also derive dynamical equations for the fields $\phi$ and $\sigma$ and discuss the topology-dependent exponents describing the coarsening of denaturation bubbles during DNA melting.   

\paragraph{Melting Curves and Phase Diagram --} We first investigate the melting behaviour of DNA by means of BD simulations of the model proposed in~\cite{Fosado2016}. The dsDNA is made up by two single-stranded chains of ``patchy-beads'' connected by permanent FENE bonds. 
Every patch-bead complex represents one nucleotide and complementary strands are paired by bonds connecting patches. We model these bonds as breakable harmonic springs, which mimic hydrogen (H) bonds between nucleotides (see SM for further details). For simplicity, we consider a homopolynucleotide (no sequence dependence)~\cite{Gagua1981}.

As anticipated, the double-helical structure can be opened up \emph{in vitro} either by increasing the temperature ($T$), or by increasing pH or salt concentration: both methods effectively reduce the strength of the H bonds, $\epsilon_{\rm HB}$, in between nucleotides. The simulations reported in this Letter emulate the latter route: starting from an equilibrated dsDNA molecule, we perform a sudden quench of $\epsilon_{\rm HB}$, 
and record the time evolution of the system until a new steady state is reached (see SM Fig.~S\ref{fig:fvst}).

\begin{figure}[t]
	\includegraphics[width=0.49\textwidth]{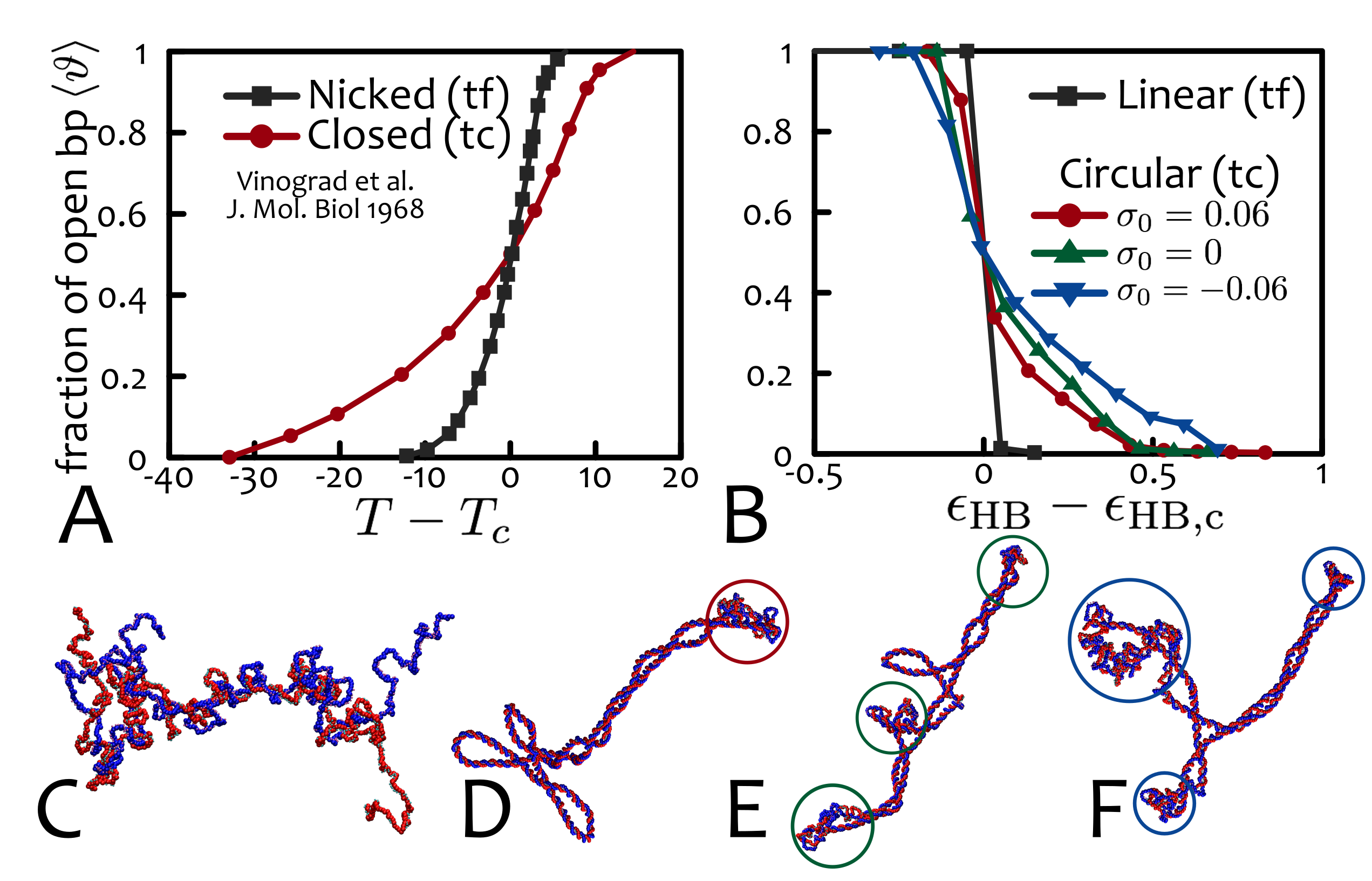}
	\vspace*{-0.3 cm}
	\caption{\textbf{Melting curves.} (\textbf{A}) shows the melting curves (fraction of denatured bp, $\langle \vartheta \rangle$) for nicked and intact polyoma DNA as a function of $T-T_c$ (data from~\cite{Vinograd1968}). (\textbf{B}) shows the melting curves obtained in the present work via BD simulations of tf and tcDNA molecules, with length 1000 bp and different levels of supercoiling, as a function of the (shifted) effective hydrogen-bond strength $\epsilon_{\rm HB}-\epsilon_{\rm HB,c}$ (averaged over 5 replicas and $10^6$ BD timesteps). In both experiments (\textbf{A}) and simulations (\textbf{B}), the transition appears smoother for tcDNA and the relative broadening $\left. \Delta t \right|_{\rm tc} / \left. \Delta t \right|_{\rm tf} \sim 3$ is in quantitative agreement. The critical bond energies for which half of the base-pairs melt are $\epsilon_{\rm HB,c}/k_B T=1.35$ for linear DNA and $0.309,0.238,0.168$ for $\sigma_0=-0.06, 0$ and $0.06$, respectively. From these values one can readily notice that the critical bond energy decreases (linearly) with supercoiling. (\textbf{C})-(\textbf{F}) show snapshots of typical configurations for $\epsilon_{\rm HB}=0.3$ $k_BT$ for tf (linear) and tcDNA with $\sigma_0=0.06,0,-0.06$, respectively. Stably denatured bubbles localise at regions of high curvature (tips of plectonemes~\cite{Matek2014}, highlighted by circles). In (\textbf{C}) the linear DNA molecule is in a fully denatured state.}
	\vspace*{-0.5 cm}
	\label{fig:phasetransition}
\end{figure}

An observable that directly compares with  
experiments is the fraction of denatured base-pairs (bp), $\vartheta$. The plot of the equilibrium value $\langle \vartheta \rangle$ as a function of temperature or bond strength can be identified with the melting curve for DNA. Typical profiles obtained from experiments~\cite{Vinograd1968} and BD simulations performed in this work, are shown in Fig.~\ref{fig:phasetransition}(A-B): the qualitative agreement is remarkable.  

The sharpness of the melting transition can be quantified in terms of the maximum value attained by the differential melting curve as 
$\Delta t = \left| d\langle \vartheta \rangle / dt \right|^{-1}$, 
where $t$ can either be temperature, $T$, or effective hydrogen bond strength, $\epsilon_{\rm HB}$, depending on the denaturation protocol. Quantitatively, Figures~\ref{fig:phasetransition}(A)-(B) show that experiments and simulations agree in predicting melting curves for tcDNA about three times broader than for tfDNA, i.e. $\left. \Delta t \right|_{\rm tc} / \left. \Delta t \right|_{\rm tf} \simeq 3$.  
 
From these observations, it is clear that the melting behaviour of DNA is affected by global topology. On the other hand, melting occurs through local opening of the double-helical structure. The challenge faced by a theory aiming to understand the ``helix-coil'' transition in tcDNA is therefore to capture local effects due to the global topological invariance. To this end, it is useful to define an effective local supercoiling field $\sigma(x,t) \equiv  \left( Lk(x,t) - Lk_0\right)/Lk_0$, where $Lk_0$ is the linking number between the two strands in the relaxed B-form state, i.e. 1 every 10.4 bp, and $Lk(x,t)$ is the effective linking number at position $x$ and time $t$.  
For a circular closed molecule of length $L$ 
\begin{equation}
\dfrac{1}{L}\int_0^L \sigma(x,t) dx = \sigma_0 \, \forall t \, ,
\label{eq:scoil}
\end{equation}
where $\sigma_0$ is the initial supercoiling deficit, which can be introduced and locked into the chain by, for instance, the action of topological enzymes~\cite{Alberts2014}. On the contrary, circular nicked or linear (tf) dsDNA molecules need not satisfy Eq.~\eqref{eq:scoil}, since any deviation from the relaxed supercoiling state can be expelled through the chain ends or the nick.
In light of this, it is clear that subjecting a tcDNA to denaturation-promoting factors causes a competition between entropy and torsional stress: the former associated with the denatured coiled regions~\cite{Poland1966a,Kafri2000}, the latter arising in the intact helical segments~\cite{Kabakcioglu2009a}. 


Motivated by these observations, we propose the following phenomenological mean field theory for the melting of tf and tcDNA. We consider a denaturation field, $\phi(x,t)$, describing the state of base-pair $x$ at time $t$ (e.g., taking the value $0$ if intact or $>0$ if denatured), coupled to a conserved field, $\sigma(x,t)$, tracking the local supercoiling. A Landau free energy can be constructed by noticing that: (i) the denaturation field $\phi$ should undergo a first-order phase transition when decoupled from $\sigma$~\cite{Poland1966a}, (ii) the elastic response to the torsional stress should be associated with even powers of $\sigma$~\cite{Benham1979}~\footnote{This is a good approximation for small $\sigma$.},  and (iii) the coupling between $\sigma$ and $\phi$ should be such that there should be an intact dsDNA phase at sufficiently low $T$,  i.e. $\phi=0$ for any $\sigma_0$ at $T<T_c$. 

Based on these considerations we can write an effective free energy density as: 
\begin{align}\label{eq:fe}
	\beta f(\phi,\sigma)  =& \left(\dfrac{b^2}{4 c} + 1-a(T)\right)\phi^2 + b \phi^3 + c \phi^4 + \notag \\
	 & + a_\sigma \sigma^2 + b_\sigma \sigma^4 + \chi \sigma \phi^2 \, .
\end{align}
In Eq.~\eqref{eq:fe}, the first term is written so that the parameter $a(T) \sim T/T_c$, $b<0$ and
we keep the quartic term in $\sigma$ to ensure there is a local minimum around $\sigma=-1$ (or $Lk(x,t)=0$) when $\phi=\phi_0$, corresponding to the fully denatured state (see SM). 
Finally, the coupling term $\chi \sigma \phi^2$ models the interplay between supercoiling and local melting; it can be turned off by simply setting $\chi=0$ to approximate tfDNA 
(in this framework the torsional stress can be expelled infinitely fast). We also highlight that by setting $\chi=0$, Eq.~\eqref{eq:fe} predicts a first-order melting transition.



The free energy density in Eq.~\eqref{eq:fe} displays a minimum at $\phi=0=\phi_{\rm ds}$ (helical state), and can develop a competing one at $\phi=\phi_{\rm ss}>1$ (coiled state) which in general depends on $\sigma$, $\chi$ and $a$ (see SM). The free energy density of the two becomes equal at the critical temperature $a=a_c$, which reduces to $a_c(\sigma)=1+\chi \sigma$, by fixing $c=-b/2$ (see SM). 
This relation states that more negatively supercoiled molecules denature at lower temperatures, as in experiments~\cite{Viglasky2000}, whereas for tfDNA ($\chi=0$) the critical temperature is independent on supercoiling.

\begin{figure}[t]
	\includegraphics[width=0.45\textwidth]{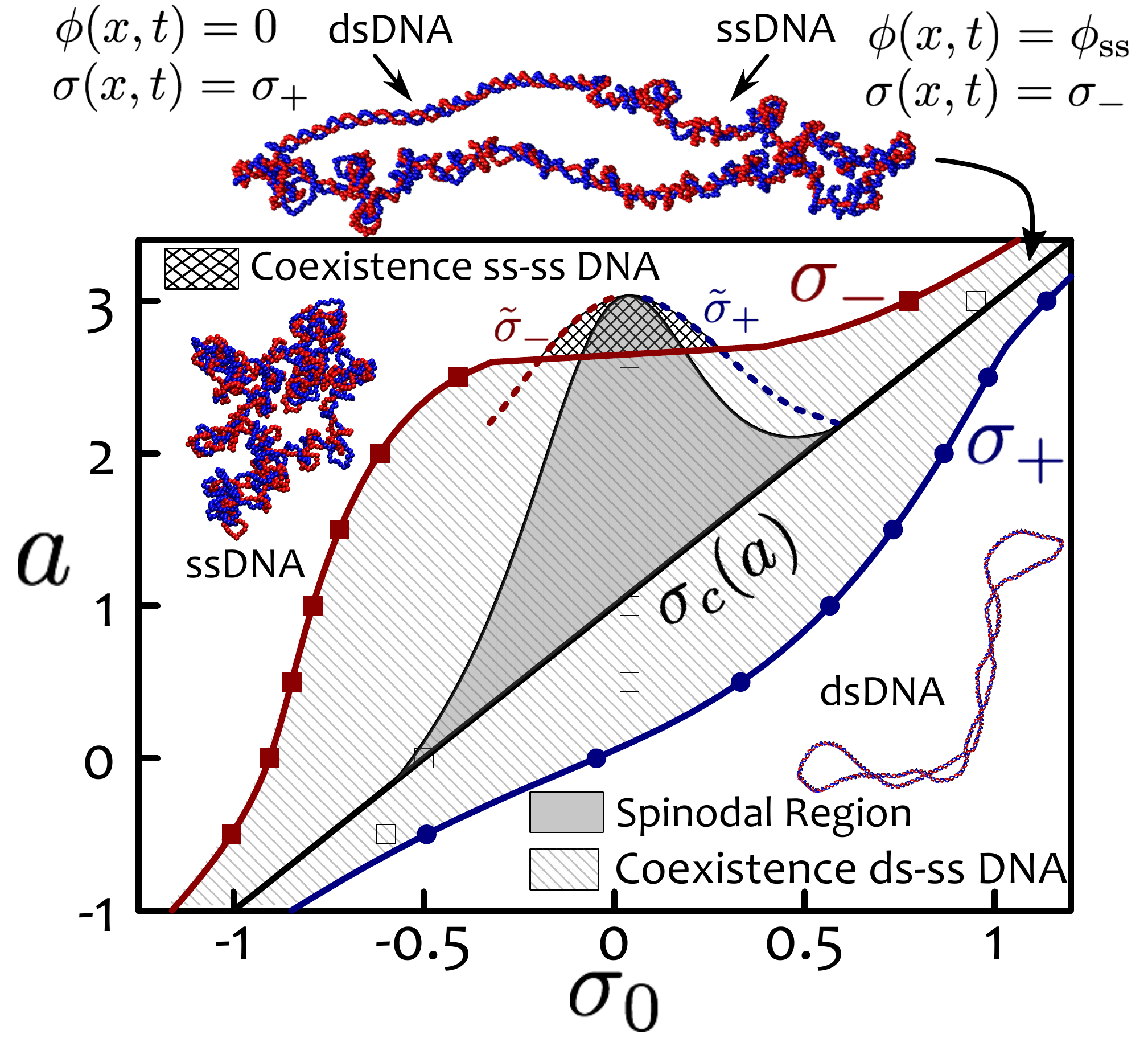}
	\vspace*{-0.5 cm}	
	\caption{\textbf{Phase Diagram.} The thick solid line represents the hidden first order transition line $\sigma_c(a)$. The line-shadowed area highlights the region of absolute instability of the uniform phase; the spinodal region is coloured in grey. Binodal lines are denoted as $\sigma_-$ and $\sigma_+$. Cross-shadowed area highlights the region of coexistence of two denatured (ss) phases. Filled symbols denote the values obtained from numerical integration of Eq.~\eqref{eq:modelc}, with initial $\sigma_0$ as indicated by the empty squares. Snapshots of ssDNA, dsDNA and ds-ss DNA coexistence observed in BD simulations are also shown.}
	\label{fig:phasediag}
	\vspace*{-0.5 cm}
\end{figure}

To obtain the phase diagram of the system in the space $(a,\sigma_0)$ we focus on dsDNA molecules with fixed, and initially uniform, value of supercoiling $\sigma_0$, at fixed temperature $a$. For such conditions, the system attains its free-energy minimising state for a value of $\phi=\phi_0(\chi,\sigma_0,a)$~\cite{Matsuyama2002}. The uniform solution $(\sigma_0,\phi_0)$ is linearly unstable if it lies within the spinodal region (in Fig.~\ref{fig:phasediag} shaded in grey), 
i.e. where $\partial^2 f(\phi_0,\sigma) / \partial \sigma^2\le 0$~\cite{ChaikinLubensky}. In Figure~\ref{fig:phasediag} we fix for concreteness  $b=-4$, $a_\sigma=1$, $b_\sigma=1/2$, $\chi=2$ (different parameter choices lead to similar diagrams provided $b$ remains negative~\footnote{Given these values of $b$, $a_{\sigma}$, $b_{\sigma}$, we further require $1.72 \leq \chi \leq 2.67$ to obtain a spinodal region.}).

A system with unstable uniform solution separates into two phases with low ($\sigma_-$) and high ($\sigma_+$) supercoiling levels, as this lowers the overall free energy. The values of $\sigma_-(a)$ and $\sigma_+(a)$ are the coexistence curves, or binodals, which are found by imposing that both chemical potential $\mu(s) \equiv \left. \partial f(\phi_0,\sigma)/\partial \sigma \right|_s$ and pressure $\Pi(s)= f(\phi_0,s)- \mu(s) s$ must be equal in the two phases~\cite{Matsuyama2002}. This translates into solving a system of two equations with two unknowns,   
\begin{align}
&\mu(\sigma_-)=\mu(\sigma_+) \\ 
&\Pi(\sigma_-)=\Pi(\sigma_+).
\end{align}
By noticing that $\sigma_0$ needs to satisfy Eq.~\eqref{eq:scoil} for tcDNA, it is straightforward to find the fractions of the system in the high and low supercoiling phases as $f_+ = (\sigma_0 - \sigma_-)/(\sigma_+ - \sigma_-)$ and $f_- = (\sigma_+ -\sigma_0)/(\sigma_+ - \sigma_-)$, respectively.

The phase diagram in the $(a,\sigma_0)$ space is reported in Figure~\ref{fig:phasediag}, where we show that the coexistence lines $\sigma_-(a)$ and $\sigma_+(a)$ wrap around the critical first-order transition line $\sigma_c(a)=(a-1)/\chi$ which therefore becomes ``hidden''~\cite{Matsuyama2002}. 
In light of this we argue that the smoother transition observed for tcDNA~\cite{Vinograd1968,Gagua1981} can be understood as a consequence of the emergence of a coexistence region in the phase space which blurs the underlying first-order transition. This argument also explains the ``early melting'' of closed circular DNA~\cite{Vinograd1968}, which can be understood as the entry into the coexistence region from low temperatures. 

Intriguingly, our phase diagram includes a region (cross-shadowed in Fig.~\ref{fig:phasediag}) where the system displays stable coexistence of two open phases (i.e. $\phi=\phi_{\rm ss}$ in both sub-systems) with supercoiling levels $\tilde{\sigma}_-$ and $\tilde{\sigma}_+$. 

\paragraph{Dynamical Scaling --} The dynamics of the non-conserved order parameter, $\phi$, and the conserved one, $\sigma$, can be found following the Glauber and Cahn-Hilliard prescriptions, respectively. Consequently, the system can be described by the following ``model C'' equations~\cite{ChaikinLubensky}
\begin{align}
&\dfrac{\partial \phi(x,t)}{\partial t}= - \Gamma_\phi \dfrac{\delta \mathcal{H}(\phi,\sigma)}{\delta \phi}\notag \\ 
&\dfrac{\partial \sigma(x,t)}{\partial t}= \Gamma_\sigma \nabla^2 \dfrac{\delta \mathcal{H}(\phi,\sigma)}{\delta \sigma} \label{eq:modelc_general}
\end{align}
where $\Gamma_{\sigma,\phi}$ are relaxation constants, 
\begin{equation}
\mathcal{H}(\phi,\sigma)=\int \left( f(\phi,\sigma)+ \gamma_\phi \left(\nabla \phi\right)^2 + \gamma_\sigma \left(\nabla \sigma\right)^2 \right) dx \, , 
\end{equation} 
and $\gamma_{\phi,\sigma}$ determine the effective surface tension of bubbles and supercoiling domains, respectively.
From Eqs.~\eqref{eq:fe} and \eqref{eq:modelc_general} one can write
\begin{align}
&\dfrac{\partial \phi(x,t)}{\partial t}= \notag\\
&=- \Gamma_\phi \left[  2\left(\dfrac{b^2}{4 c} + 1 - a \right)\phi + 3 b \phi^2 + 4 c \phi^3 + 2 \chi \phi \sigma - \gamma_\phi \nabla^2 \phi \right] \notag \\ 
&\dfrac{\partial \sigma(x,t)}{\partial t}= \Gamma_\sigma \nabla^2 \left[ 2 a_\sigma \sigma + 4 b_\sigma \sigma^3 + \chi \phi^2 - \gamma_\sigma \nabla^2 \sigma \right] \, . \label{eq:modelc}
\end{align}
We numerically solve this set of partial differential equations (PDE) on a 1D lattice of size $L$ for fixed $a$ and $\sigma_0$ (see SM for details) and compare the evolution of denaturation bubbles with the one observed in BD simulations. Note this set of equations disregards thermal noise, hence it is in practice a mean field theory.

\begin{figure}[t]
	\centering
	\includegraphics[width=0.49\textwidth]{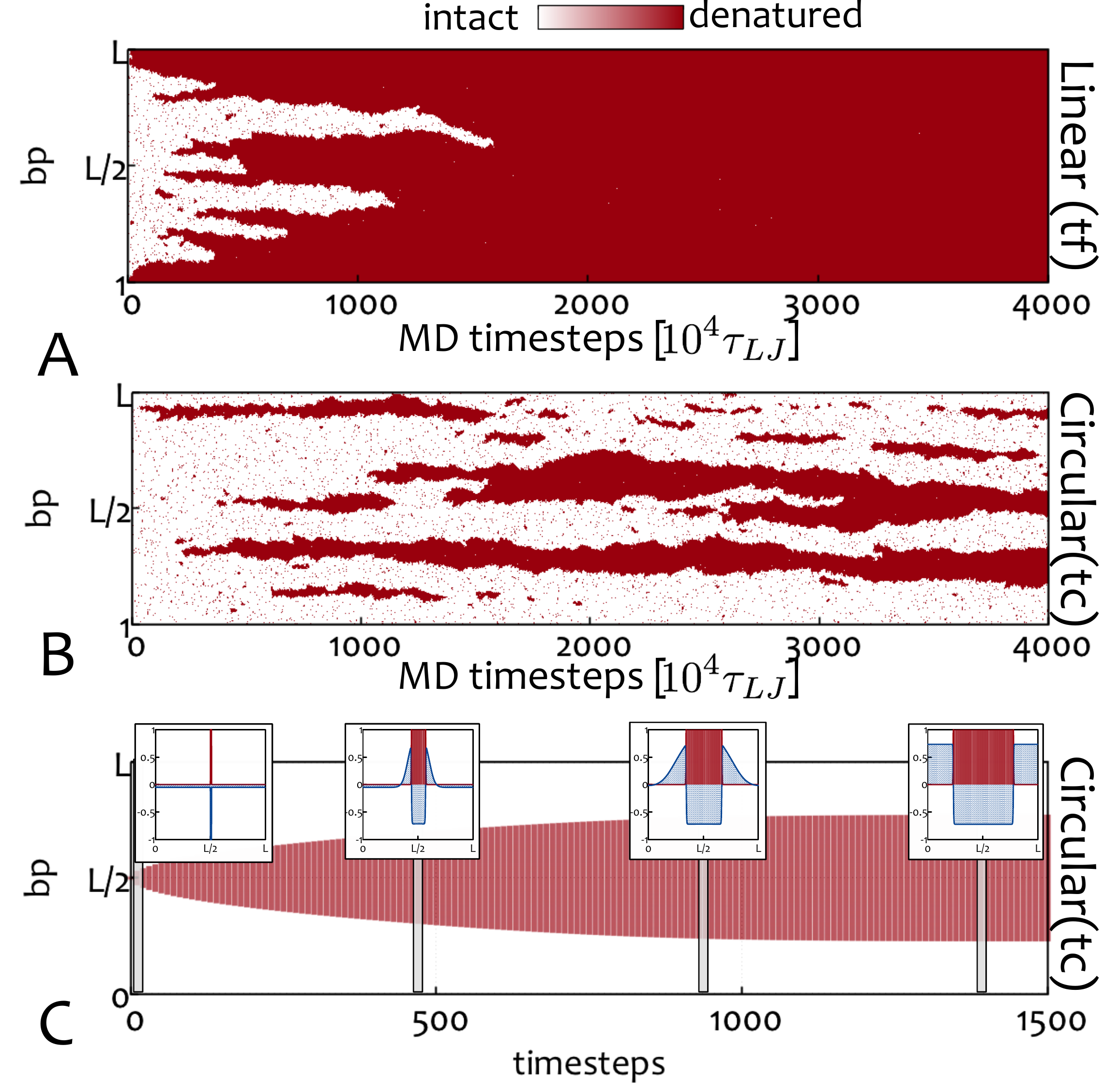}
	\caption{\textbf{Kymographs.} (\textbf{A}-\textbf{B}) report results from BD simulations. At time $t=0$, the H bond strength is quenched to $\epsilon_{\rm HB} = 0.3 k_BT$ and the local state of the chain (red for denatured and white for intact) is recorded as a ``kymograph''. (\textbf{A}) and (\textbf{B}) show the case of a tf and tcDNA  ($\sigma_0=0$), respectively. 
	(\textbf{C}) shows the kymograph of the system during integration of Eqs.~\eqref{eq:modelc} starting from a small bubble (see SM). Insets show instantaneous profiles of denaturation field (red) and supercoiling field (blue).}
\vspace*{-0.5 cm}
	\label{fig:kymo}
\end{figure}

In Figure~\ref{fig:kymo} we show ``kymographs'' from BD simulations, capturing the state of each base-pair (either intact or denatured) against time for tf and tcDNA. As one can notice, after the energy quench at $t=0$, the linear (tfDNA) molecule starts to denature from the ends and eventually fully melts. On the other hand, in the closed circular (tcDNA) molecule, bubbles pop up randomly over the whole contour length, and the steady state entails a stable fraction, $0<\vartheta<1$, of denatured bp (see also SM, Fig.~S\ref{fig:fvst}). 

We observe a similar behaviour when the fields $\phi$ and $\sigma$ are evolved via Eqs.~\eqref{eq:modelc}, starting from a single small bubble at temperature $a$ within the coexistence region (see Fig.~\ref{fig:kymo}D and SM). While the bubble grows, the supercoiling field is forced outside the denatured regions and accumulates in the ds segments. The increasing positive supercoiling in the helical domains slows down and finally arrests denaturation, resulting in phase coexistence in steady state, between a denatured phase with $\sigma=\sigma_-$ and $\phi=\phi_{\rm ss}>1$, and an intact phase with $\sigma=\sigma_+$ and $\phi=\phi_{\rm ds}=0$. 

\begin{figure}[b]
	\centering
	\includegraphics[width=0.49\textwidth]{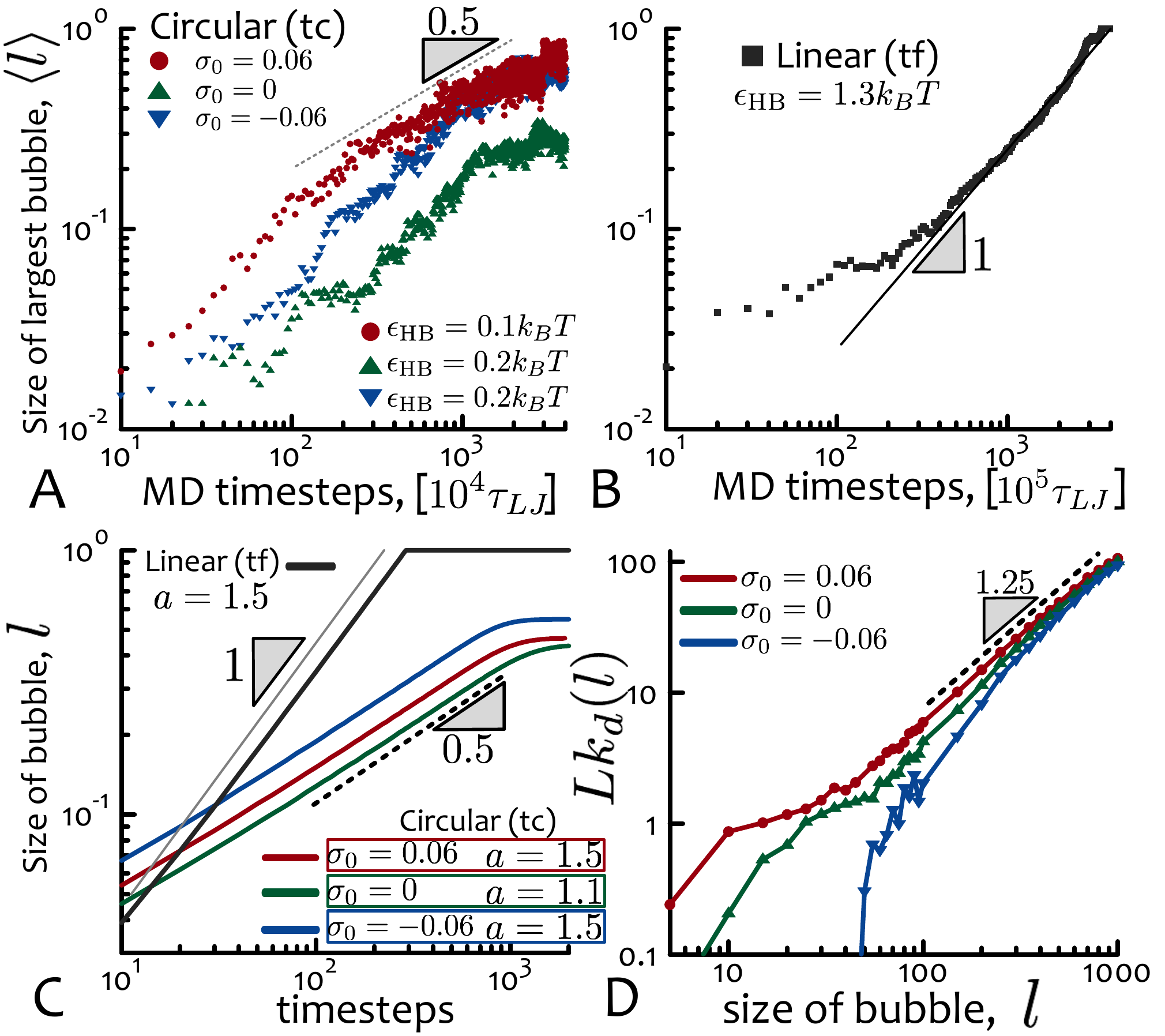}
	\caption{\textbf{Dynamical Scaling.} (\textbf{A}-\textbf{B}) show results from BD simulations. The size of the largest denatured bubble $\langle l \rangle$ (averaged over 5 replicas) is plotted against time from the moment of the quench. (\textbf{A}) shows tcDNA while (\textbf{B}) refers to tfDNA. (\textbf{C}) shows the size of a single growing bubble, $l$, within our mean field model, Eqs.~\eqref{eq:modelc}. PDE and BD simulations show similar behaviours, which suggest a universal dynamical scaling with topology-dependent exponent ($\alpha=1$ for $\chi=0$ and $\alpha=1/2$ for $\chi>0$). (\textbf{D}) shows the linking number, $Lk_d$, stored inside a denatured bubble of fixed size $l$ computed from BD simulations (see text and SM for details).}
	\label{fig:scaling}
\end{figure}

The growth, or coarsening, of a denaturation bubble, $l$, can be quantified within our mean field theory and BD simulations: in the former case by numerical integration of Eq.~\eqref{eq:modelc}, in the latter by measuring the size of the largest bubble over time and averaging over independent realisations. 
As shown in Fig.~\ref{fig:scaling}(A-C) we find that in both mean field and BD simulations, 
\begin{equation}
l(t) \sim 
\begin{cases}
t^1 \hspace{0.8 cm} \text{ for topologically free DNA,} \\
t^{1/2} \hspace{0.5 cm} \text{ for topologically constrained DNA.}
\end{cases}
\end{equation}
In other words, we find that the exponent $\alpha$ governing the \emph{local} growth of a denaturation bubble depends on the \emph{global} topology of the molecule. 


We propose the following argument to explain the values of $\alpha$. For tfDNA (e.g., nicked or linear), we can assume that the supercoiling field relaxes quickly, and gets expelled outside, without affecting the dynamics of the denaturation field. In this case, the free energy can be approximated as $f \simeq (\epsilon_{\rm HB} - T \Delta S)l$, so that there is a constant increase in entropy per each denatured bp when $T > T_c=\epsilon_{\rm HB}/\Delta S$. This implies that~\cite{ChaikinLubensky} $\psi dl/dt \simeq  df/dl \sim const$, with $\psi$ an effective constant friction; as a result we obtain $l(t)\sim t$.

On the other hand, the value of $\alpha=1/2$ observed for tcDNA (e.g., circular non-nicked plasmids) can be understood by quantifying the slowing down of denaturation due to the accumulation of a ``wave'' of supercoiling, raked up on either side of the growing bubble. We argue that the flux of $\phi$ through a base pair at the bubble/helix interface is $J_\phi \sim \phi_{\rm ss} dl/dt$. At the same time, the flux of $\sigma$ can be obtained by noticing that the ``wave'' can be approximated by a triangle with constant height $h=\sigma_+-\sigma_0$ and base $b \sim l(t)$ (see SM Fig.~S\ref{fig:fieldprofile} and Fig.~\ref{fig:kymo}(C)). This is because the total supercoiling enclosed by the wave must be proportional to the one expelled from within the denatured bubble, which is $\sim \left| \sigma_- - \sigma_0 \right| l(t)$. 
One can therefore write	that $J_\sigma=-\Gamma_\sigma \partial_x \sigma  \simeq \Gamma_\sigma h/l$, for the supercoiling flux in, say, the forward direction. At equilibrium, the two fluxes must balance, i.e. $J_\phi \sim J_\sigma$, and therefore $\phi_{\rm ss}dl/dt \sim \Gamma_\sigma h/l$, or $l(t) \sim t^{1/2}$. Finally, we highlight that this argument depends on the slowing down over time of 1D supercoiling fluxes, hence is qualitatively distinct from the reason why $\alpha=1/2$ in dimensions $d\geq 2$ in (non-conserved) model A~\cite{ChaikinLubensky}.

\paragraph{Linking within Denaturation Bubbles --}
As a final result, we perform BD simulations to characterise the topology of a denaturation bubble through the linking number that can be stored inside it (Fig.~\ref{fig:scaling}(D)). An idealised bubble is identified by ${\rm Lk}=0$ (and $\sigma=-1$)~\cite{Kabakcioglu2009a}, whereas our BD simulations show that a denatured region of {\it fixed} size $l$ (imposed by selectively breaking \emph{only} $l$ consecutive bonds along an intact dsDNA molecule)
has a non-zero linking number $Lk_d$ (see Fig.~\ref{fig:scaling}(D))~\footnote{In practice, $Lk_d$ was obtained by removing the linking number of the chain \emph{outside} the denatured region $Lk_{\rm out}$ from the initially set $Lk$, see SM for details.}.

We found that for small $l$, $Lk_d$ displays a remarkable signature of global topology (through the value of $\sigma_0$); 
instead, the scaling behaviour at large $l$ appears to follow $Lk_d \sim l^{1.25}$ irrespectively of $\sigma_0$, until it reaches $Lk_0$. The finding that a denaturation bubble displays a non-trivial and $l$-dependent linking number suggests that idealised ($Lk_d=0$) bubbles may not always be reflecting realistic behaviour. Further, it may be of relevance for processes such as DNA replication, as it suggests that supercoiling or torsional stress may be able to diffuse past branching points such as replication forks~\cite{Postow2001}.

\paragraph{Conclusions --} In summary, we have studied the melting behaviour of topologically constrained DNA through a combination of large-scale BD simulations and mean field theory. A key result is that the phase diagram for tcDNA melting generally involves a phase coexistence region between a denatured and an intact phases, pre-empting 
a first-order denaturation transition as in tfDNA. This finding provides a theoretical framework to explain the long-standing experimental observation that the denaturation transition in circular, and not nicked, supercoiled plasmids is seemingly less cooperative (smoother) than for linear, or nicked, DNA~\cite{Vinograd1968,Gagua1981}. 

We have further studied, for the first time, the coarsening dynamics of denaturation bubbles in tcDNA, and found a remarkable agreement between BD simulations and mean field theory, both reproducing similar topology-dependent scaling exponents that can be understood within our theoretical model. It would be of interest to investigate such dynamics experimentally in the future.

DMa and DMi acknowledge ERC for funding (Consolidator Grant THREEDCELLPHYSICS, Ref. 648050). YAGF acknowledges support form CONACyT PhD grant 384582. 

\vspace*{-0.5 cm}

\bibliographystyle{apsrev4-1}
\bibliography{Denaturation,dsDNA,lammps}

\clearpage

\setcounter{section}{0}
\setcounter{figure}{0}
\setcounter{table}{0}
\setcounter{equation}{0}

\renewcommand{\figurename}{Fig.~S}
\renewcommand{\tablename}{Table S}

{\large \bf Supplementary Material}

\tableofcontents

\clearpage

\section{Brownian Dynamics Simulations of double stranded DNA}\label{sec:ModDetails}
Brownian Dynamics simulations are performed using the model developed in Ref.~\cite{Fosado2016} to which we refer for details on parameters and validation. Below we briefly describe its main features. \\

\begin{figure}[t]
	\includegraphics[width=0.45\textwidth]{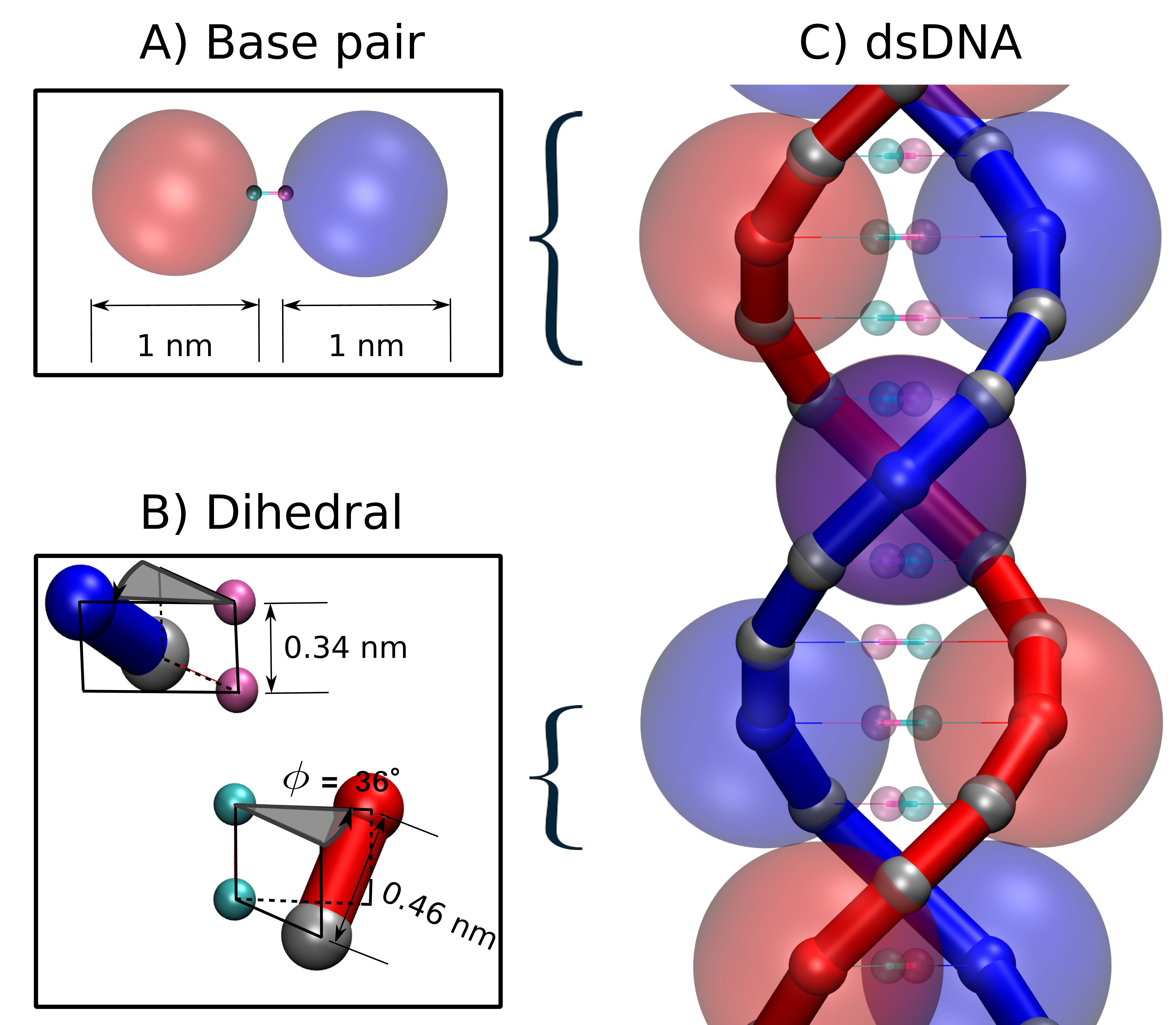}
	\caption{\textbf{Model.} (A) A base pair is formed by two nucleotides bonded via a breakable harmonic spring. The parameters are set to reproduce the thickness of B-DNA, around $2$ nm. (B) The equilibrium distance between consecutive beads in the same strand is set to $0.46$ nm, while the distance between its corresponding patches is $0.34$ nm. This choice of the parameters imposes a twist of $\phi = 36^{\circ}$ between consecutive base-pairs. Additionally, the handedness of the chain is imposed with a dihedral interaction. (C) Diagram of the dsDNA structure, where shaded beads show the excluded volume of the interacting beads in the two strands (blue and red). The grey spheres represent the non-sterically interacting beads.}
	\label{fig:model}
\end{figure}

The double-stranded (ds) DNA molecule is modelled as two polymers made of ``patchy-beads'' connected by springs. Specifically, each nucleotide is represented by a rigid body made up of two spherical monomers: a bead which models the sugar-phosphate backbone, and a ``patch'' that represents the nitrogenous base. Two nucleotides belonging to complementary strands are shown in Figure~S\ref{fig:model}(a). The beads have an excluded volume of $\sigma = 1$ nm in diameter and they are shown in red for one strand and in blue for the complementary. The corresponding patches are shown in cyan and pink: these have no associated excluded volume and are placed at a distance of $\sigma/2$ from the centre of the bead. Excluded volume between beads is modelled via a truncated and shifted Lennard-Jones (LJ) potential (also known as Weeks-Chandler-Anderson potential),
\begin{equation}
U_{LJ}(r)  =  4\epsilon \left[ 
\left( \frac{\sigma}{r} \right)^{12} 
- \left( \frac{\sigma}{r} \right)^{6} + \frac{1}{4} \right] \, ,
\end{equation}
if $r<2^{1/6}\sigma$, and $U_{LJ}(r)=0$ otherwise. 

The hydrogen bonding interaction, responsible for holding two nucleotides belonging to complementary strands together, is modelled by \emph{breakable} harmonic springs acting between two patches (Fig.~S\ref{fig:model}(a)). The potential associated to this interaction is
\begin{equation}
U_{\rm hb}(r) =  \dfrac{\epsilon_{HB}}{(r_{\rm 0, hb}-r_{\rm c, hb})^{2}}\left[ (r-r_{\rm 0, hb})^{2}-(r_{\rm c, hb}-r_{\rm 0, hb})^{2}  \right]  
\end{equation}
if $r \leq r_{\rm c, hb}$ and $U_{\rm hb}=0$ otherwise. The equilibrium bond distance $r_{\rm 0, hb}$ is set to zero and the critical distance $r_{\rm c, hb} = 0.3$ nm marks the point at which the bond breaks.  The strength of the potential is $\epsilon_{HB}$, the main parameter used in the main text to drive the melting transition.

A single strand is formed by a chain of patchy-beads connected via FENE bonds of length $d_{bp} = 0.46$ nm. This potential reads
\begin{equation}
U_{bb}(r) =  -\dfrac{\epsilon_{bb}R^{2}_{0}}{2} \log{\left[ 1-\left( \frac{r}{R_{0}} \right) ^{2} \right]} \text{ if } r<R_0 
\end{equation}
with $R_0=0.68 \sigma$ and $U_{bb}(r) = \infty$ otherwise. 

We also model the stacking of base-pair by means of a Morse potential which keeps the distance between consecutive patches along each strand around $r_{\rm 0,stack}=0.34$ nm, i.e.
\begin{equation}
U_{morse}(r) = \epsilon_{m}[1-e^{-\lambda(r-r_{\rm 0,stack})}]^{2}.
\end{equation}

While the choice of the two previous potentials ensures that the pitch of the chain is the one expected for dsDNA (around 10 bp per helical turn), a dihedral interaction (see Fig.~S\ref{fig:model}(b)) between the particles of two consecutive nucleotides in the same strand enforces right-handed helicity for the DNA molecule -- i.e., we set
\begin{equation}
U_{dihedral} = \epsilon_{d} [1+\cos{(\phi-d)}],
\end{equation}
where $d=180^\circ-36^\circ$. 

Planarity between basis is imposed through a harmonic potential preventing the angle formed between two consecutive patches and one bead (all in the same strand) to change from its equilibrium value ($\alpha_0 = 90^{\circ}$).
\begin{equation}
U_{plan} = \dfrac{\epsilon_{h}}{2}(\alpha - \alpha_{0})^{2}.
\label{harm}
\end{equation}

Finally, in order to regulate the stiffness of the chain we use a Kratky-Porod potential controlling the angle between three consecutive patches along one strand, i.e.
\begin{equation}
U_{bending} = \epsilon_{b} [1+\text{cos}(\theta)].
\end{equation}
A full description of all the potentials used in the model can be found in Ref.~\cite{Fosado2016}.

In order to avoid large overlaps between beads and to preserve the correct geometry of DNA we consider two types of beads in each strand: sterically interacting
beads (shown as small solid red spheres for one strand and blue for the other in Fig.~S\ref{fig:model}(C)) are intercalated by two non-sterically interacting beads (represented as small grey spheres). These do not interact with the beads along the same strand but they repel all beads on the complementary strand within an excluded radius $\sigma/2$. This choice ensures that only non-overlapping beads along each single strand sterically interact with one another and, at the same time, allows us to preserve the topology by forbidding the two strands crossing through one another.

The nucleotides made up by the bead-patch complex evolve as a rigid bodies undergoing Langevin dynamics, i.e. the position of each bead $\bm{r}$ is updated according to equation
\begin{equation}
	m \dfrac{d^2\bm{r}}{dt^2} = -\xi \dfrac{d \bm{r}}{dt}-\bm{\nabla}\mathcal{U} + \sqrt{2 k_B T \xi} \bm{f}  
	\label{eq:langevin}
\end{equation}
where $\mathcal{U}$ is the total potential field experienced by the bead, $m$ is the mass of the bead, $\xi$ the friction and $\bm{f}$ the white noise term with zero mean and satisfying the fluctuation dissipation theorem 
\begin{equation}
	\langle f_\alpha(t) f_\beta(s)  \rangle = \delta_{\alpha \beta} \delta(s-t)  
\end{equation}
along each Cartesian component (denoted by Greek letters). Integration of Eq.~\eqref{eq:langevin} is performed using a velocity-Verlet algorithm through the LAMMPS~\cite{Plimpton1995} code, run in BD mode.

\section{Dynamics of Melting}\label{sec:equilibration}
\begin{figure}[t]
	\includegraphics[width=0.45\textwidth]{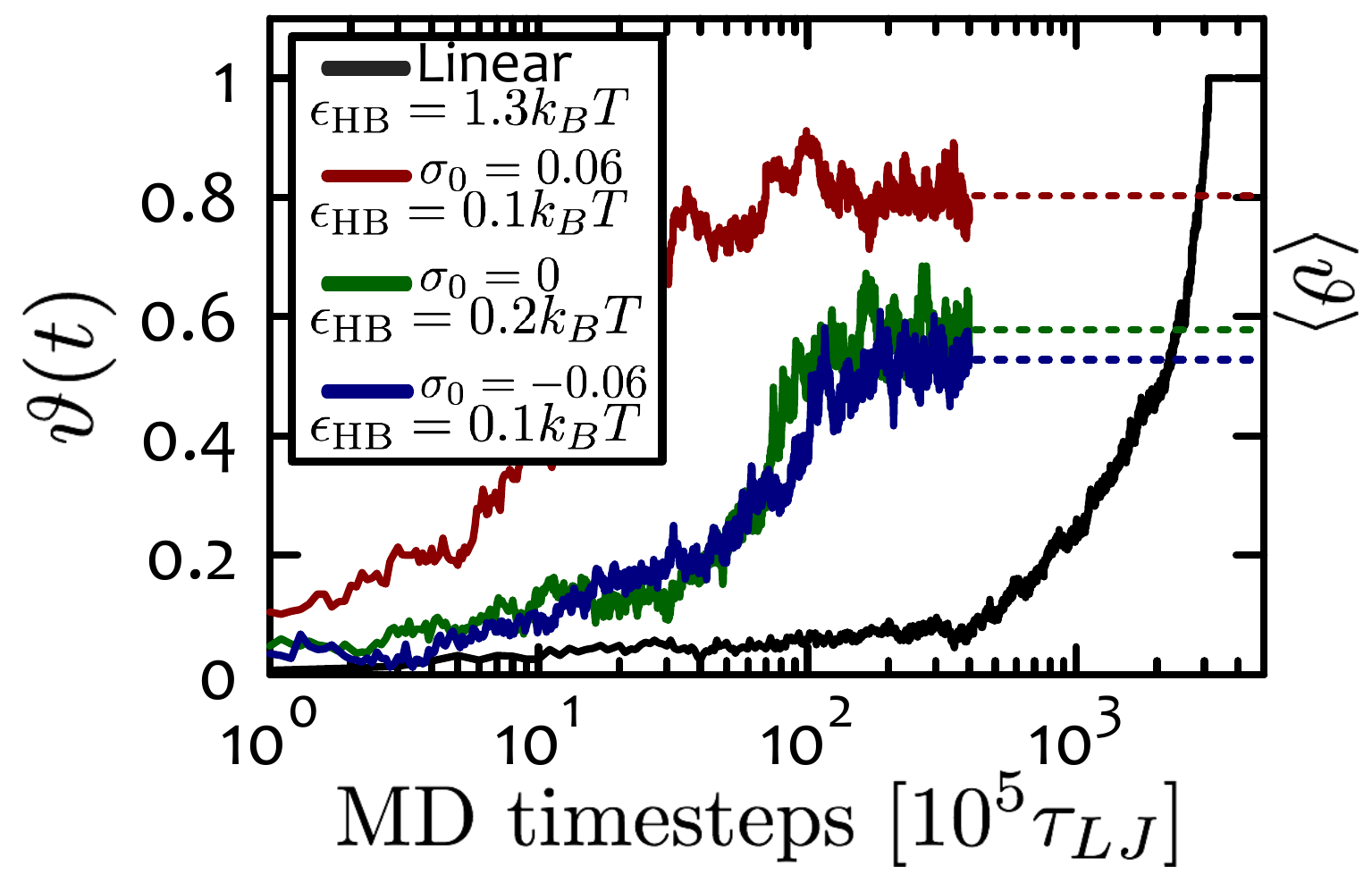}
	\caption{\textbf{Dynamics of melting.} Examples of trajectories obtained during melting of the model dsDNA in BD simulations. The curves represent the number of denatured base-pairs at a given time $t$ for linear (black line) and circular closed molecules with supercoiling $\sigma_0=0.06,0,-0.06$ plotted in red, green and blue, respectively. The values of the effective bond energy are $\epsilon_{\rm HB}=0.1,0.2$ and $0.3$, respectively. The dashed lines represent the equilibrium value $\langle \vartheta\rangle$ which give the melting curves  as a function of $\epsilon_{\rm HB}$ shown in Fig.~\ref{fig:phasetransition}. It can be seen that, while closed molecules equilibrate at a value of $\vartheta$ in between 0 and 1, the linear molecule eventually fully denatures. }
	\label{fig:fvst}
\end{figure}

In the main text we show the melting profiles $\langle \vartheta\rangle$ as a function of the effective bond strength $\epsilon_{\rm HB}$ in Figure~\ref{fig:phasetransition}. The value of stably denatured bp $\langle \vartheta\rangle$ is obtained by averaging the last $2$ $10^7$ $\tau_{LJ}$ timesteps of the trajectories obtained for $\vartheta(t)$ (examples shown in Fig.~\ref{fig:fvst}). As one can notice from the figure, all the samples start from non-denatured states $\vartheta(0)=0$; after a sudden quench in $\epsilon_{\rm HB}$ the systems evolve until a new steady state is reached. For topologically constrained (closed circular) molecules, $\vartheta(t)$ reaches a steady state value in between 0 and 1 (see red, green and blue curves). On the other hand, linear molecules initially denature very little but eventually (after a time 10 times longer than the equilibration time for rings) fully denature. 

\newpage
\phantom{A}
\newpage
\section{Landau Free-Energy}\label{sec:freeEn}
Here we study the free energy whose density is written in Eq.~\eqref{eq:fe}, 
\begin{equation}
	\beta f(\phi,\sigma) = \left(\dfrac{b^2}{4 c}+1-a(T)\right)\phi^2 + b \phi^3 + c \phi^4 + a_\sigma \sigma^2 + b_\sigma \sigma^4 + \chi \sigma \phi^2 \, . 
\end{equation}
In this equation, $\phi=\phi(x,t)$ is the field describing the state (coiled or helical) of base-pair $x$ along the contour of the DNA molecule at time $t$ and $\sigma=\sigma(x,t)$ the field quantifying the amount of local supercoiling. The parameter $a=a(T)\sim T/T_c$ is the only temperature-dependent parameter and $\chi$ the coupling between $\phi$ and $\sigma$. 

\begin{figure}[b]
	\centering
	\includegraphics[width=0.45\textwidth]{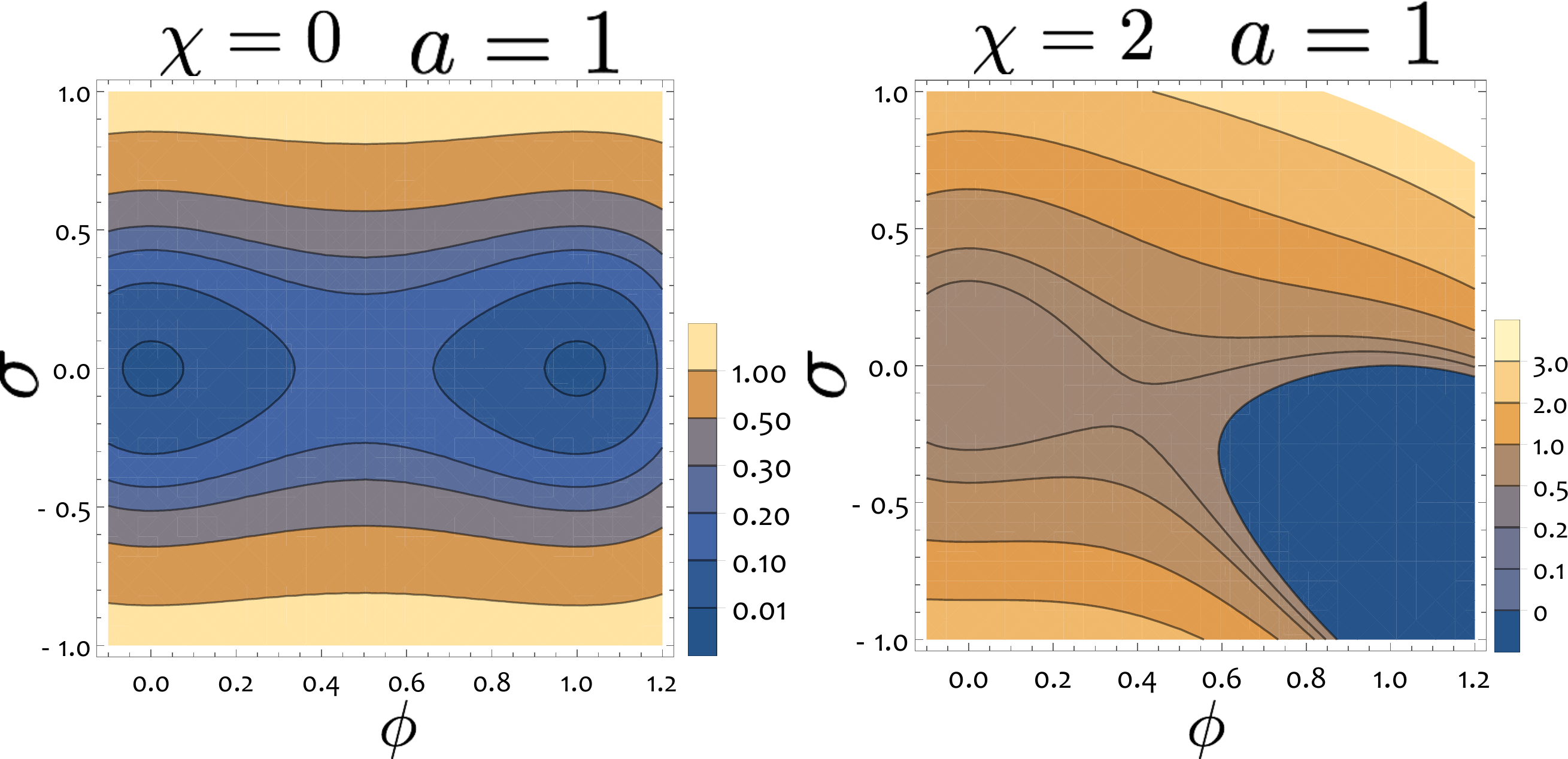}
	\caption{\textbf{Contour plot of free energy density}. Plots show $f(\phi,\sigma)$ for $a=1$ and $\chi=0$ (left) or $\chi=2$ (right).}
	\label{fig:fe_contour}
\end{figure}

This free energy density describes a first-order transition between a closed (double-stranded, ds, $\phi=\phi_{\rm ds}=0$) and an open (single stranded, ss, $\phi=\phi_{\rm ss} > 1$) DNA molecule. We choose the parameter $c=-b/2$ so that the critical temperature depends only on $\chi \sigma$ (see below) and so that the free energy density displays a minimum located at around $\sigma=-1$ when $\phi$ attains its free-energy minimising value of $\phi_0$ (see below). The corresponding choices are $c=-b/2$, $a_\sigma=1$, $b_\sigma=1/2$ and $b=-4$. The free energy density then reduces to 
\begin{equation}
	\beta f(\phi,\sigma) = \left(3-a(T)\right)\phi^2 - 4 \phi^3 + 2 \phi^4 +\sigma^2 +\dfrac{\sigma^4}{2} + \chi \sigma \phi^2 \, .
	\label{eq:fe2}
\end{equation}
As a function of $\phi$, Eq.~\ref{eq:fe2} can display two local minima whose energy is equal for topologically unconstrained ($\chi=0$) molecules at the critical (melting) temperature $T_c$ (defined via $a(T_c)=a_c=1$). In this situation (see Fig.~S\ref{fig:fe_contour}) the minima are located at $\phi_{\rm ds}=0$ (closed, double stranded) and at $\phi_{\rm ss}=1$ (open, single stranded) and mark the equal probability coexistence point.

\begin{figure*}[t]
	\centering
	\includegraphics[width=0.92\textwidth]{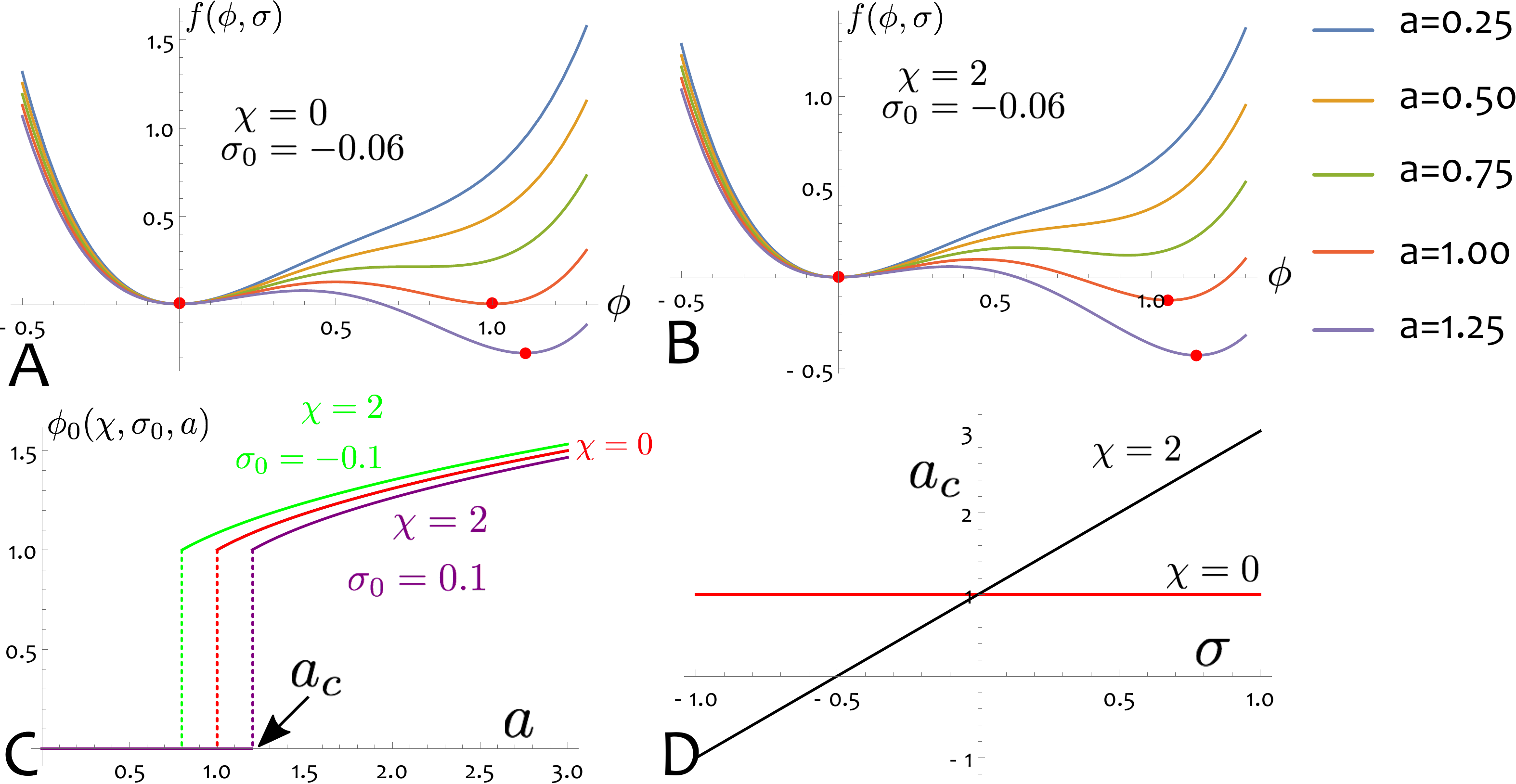}
	\caption{\textbf{First order transition for denaturation field $\phi$.} (A)-(B) Show the free energy density as a function of $\phi$. Red dots mark the location of the global minimum, i.e. $\phi=\phi_0$. (C) Shows the free-energy minimising value $\phi_0=\phi_0(\chi,\sigma_0,a)$ that discontinuously jumps at the critical temperature $a_c$. (D) Shows that the critical temperature $a_c=1$ for $\chi=0$ and $a_c \sim \sigma$ for $\chi>0$.  }
	\label{fig:fe_phi0_jump}
\end{figure*}

\begin{figure*}[t]
	\centering
	\includegraphics[width=0.92\textwidth]{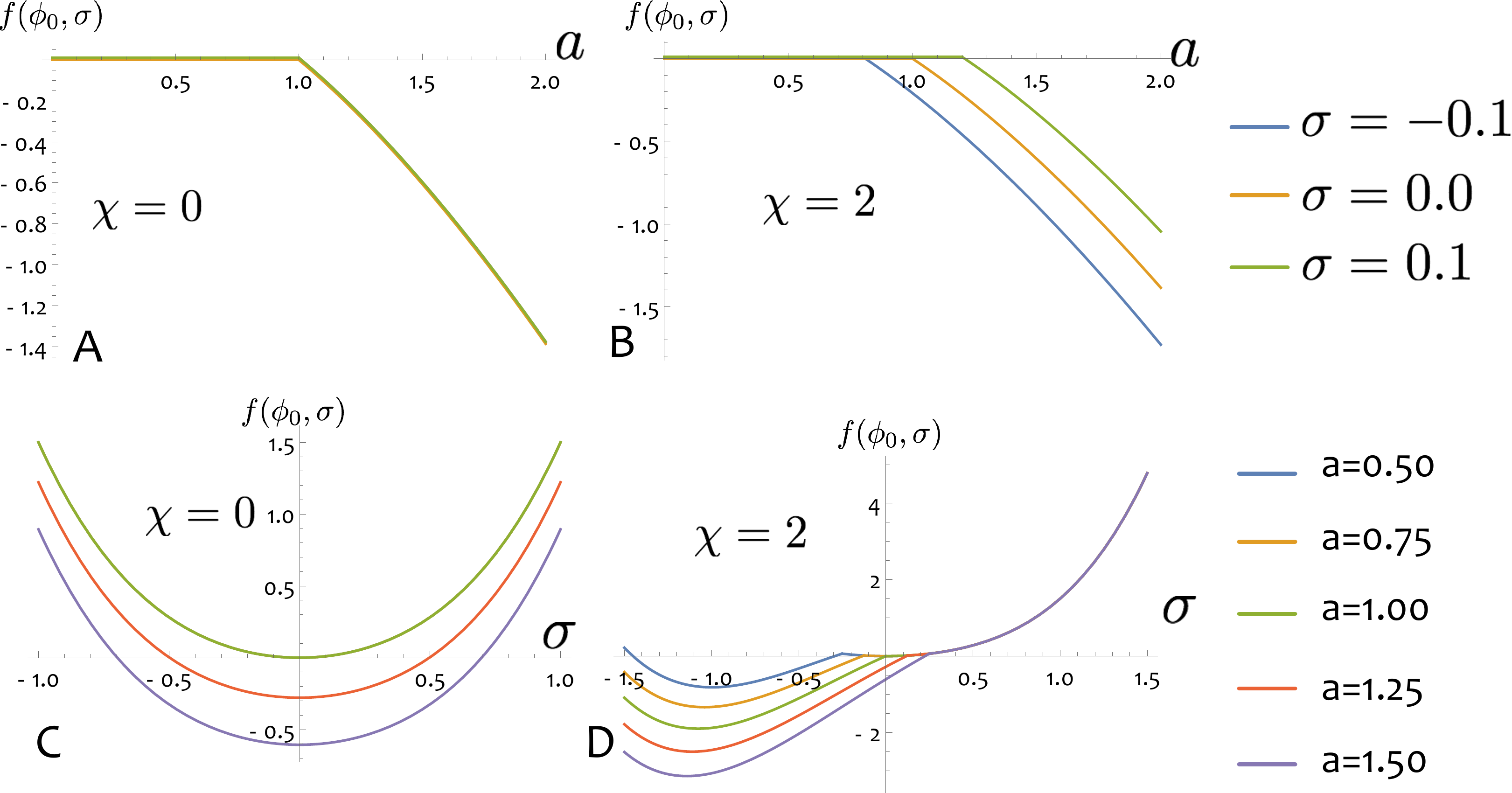}
	\caption{\textbf{Profiles of $f(\phi_0,\sigma)$.} (A) and (B) show the free energy density as a function of $a$ for $\chi=0$ and $\chi=2$.(C)-(D) Same function plotted as a function of $\sigma$ for topologically free ($\chi=0$) and topologically constrained ($\chi=2$) molecules.}
	\label{fig:fe_f_phi0}
\end{figure*}

In general, the critical temperature is supercoiling-dependent through the coupling parameter $\chi$ (see Fig.~S\ref{fig:fe_phi0_jump}A-C). The function $a_c(\sigma,\chi)$ is simply found by solving $f(\phi_{\rm ds},\sigma)=f(\phi_{\rm ss},\sigma)$, i.e. 
\begin{equation}
a_c(\sigma,\chi)=\dfrac{b^2+4c(1 + b + c)}{4 c} + \chi \sigma \, \end{equation}
or
\begin{equation}
	a_c(\sigma,\chi)=1+\chi \sigma \,
	\label{eq:ac}
\end{equation}
by substituting the values of parameters. We report $a_c$ in Fig.~S\ref{fig:fe_phi0_jump}(D) for $\chi=0$ and $\chi=2$. It can be readily appreciated that for topologically constrained ($\chi > 0$) molecules, the critical temperature $a_c$ depends linearly on the supercoiling, in agreement with experimental observations~\cite{Viglasky2000}.  

The equilibrium value of the denaturation field $\phi_0$ is a discontinuous function of the temperature and is found by minimisation of the free energy, as 
\begin{equation}
\phi_0=\min_\phi{\left\{ \dfrac{\partial f(\phi,\sigma)}{\partial \phi} = 0 \right\}}.
\end{equation}
This condition leads to (see also curves in Fig.~S\ref{fig:fe_phi0_jump}(C))
\begin{equation}
	\phi_0(\chi,\sigma,a) = 
	\begin{cases}
	0 \text{ if } a < a_c(\sigma,\chi)\\
	\dfrac{\left( 3 + \sqrt{-3 + 4 a - 4 \chi \sigma} \right)}{4} \text{ if } a \geq a_c(\sigma,\chi) \\
	\end{cases} \, .
\end{equation}  

Because $\phi$ is a non-conserved order parameter, it will always attain the value $\phi_0(\chi,\sigma,a)$ for any given set of $\chi$, $\sigma$ and $a$. The free energy density of the system can therefore be written as $f(\phi_0(\chi,\sigma,a),\sigma)=f(\phi_0,\sigma)$. The behaviour of this function is plotted in Figure~S\ref{fig:fe_f_phi0}. 

One can readily check that for $\chi=0$ the discontinuity is always located at $a_c=1$, irrespective of $\sigma$, and that the profile of the free energy is symmetric with respect to $\sigma$. On the contrary, for topologically constrained ($\chi>0$) molecules, the critical temperature is supercoiling-dependent, $a_c=a_c(\sigma)$, and the profile as a function of $\sigma$ is more complex. The term $\sigma^4$ in the free energy and the parameters chosen ensure that  $f(\phi_0,\sigma)$ displays a minimum around $\sigma=-1$ for $\phi_0=\phi_{\rm ss}$ which corresponds to a fully denatured state with no supercoiling ($Lk=0$) within the coiled region.

Figure~S\ref{fig:fe_f_phi0}(D) can be understood as follows: for fixed temperature $a$ the critical supercoiling $\sigma_c$ can be found through eq.~\eqref{eq:ac} as 
\begin{equation}
\sigma_c(a)=\left(a-1\right)/\chi \, .
\end{equation}
At $\sigma_c(a)$ the system undergoes a first-order transition (corresponding to the discontinuity of the curves in Fig.~S\ref{fig:fe_f_phi0}(D)), and the system switches equilibrium state from $\phi_{\rm ds}$ to $\phi_{\rm ss}$. This corresponds to a cusp in the behaviour of $f(\phi_0,\sigma)$. In order to find the coexisting phases we now need to perform the common tangent construction on the function $f(\phi_0,\sigma)$ represented in Figure~S\ref{fig:fe_f_phi0}(D)).\\

\newpage
\phantom{A}
\newpage
\phantom{A}
\newpage

\section{Spinodal Region and Binodal Lines}
The supercoiling field $\sigma$ is a globally conserved field, and the system may therefore undergo phase decomposition in regions of high and low $\sigma$ in order to lower the overall free energy while preserving the initially set $\sigma=\sigma_0$. The region in the parameter space $(a,\sigma_0)$ for which this occurs is know as coexistence region.

\begin{figure}[b]
	\centering
	\includegraphics[width=0.45\textwidth]{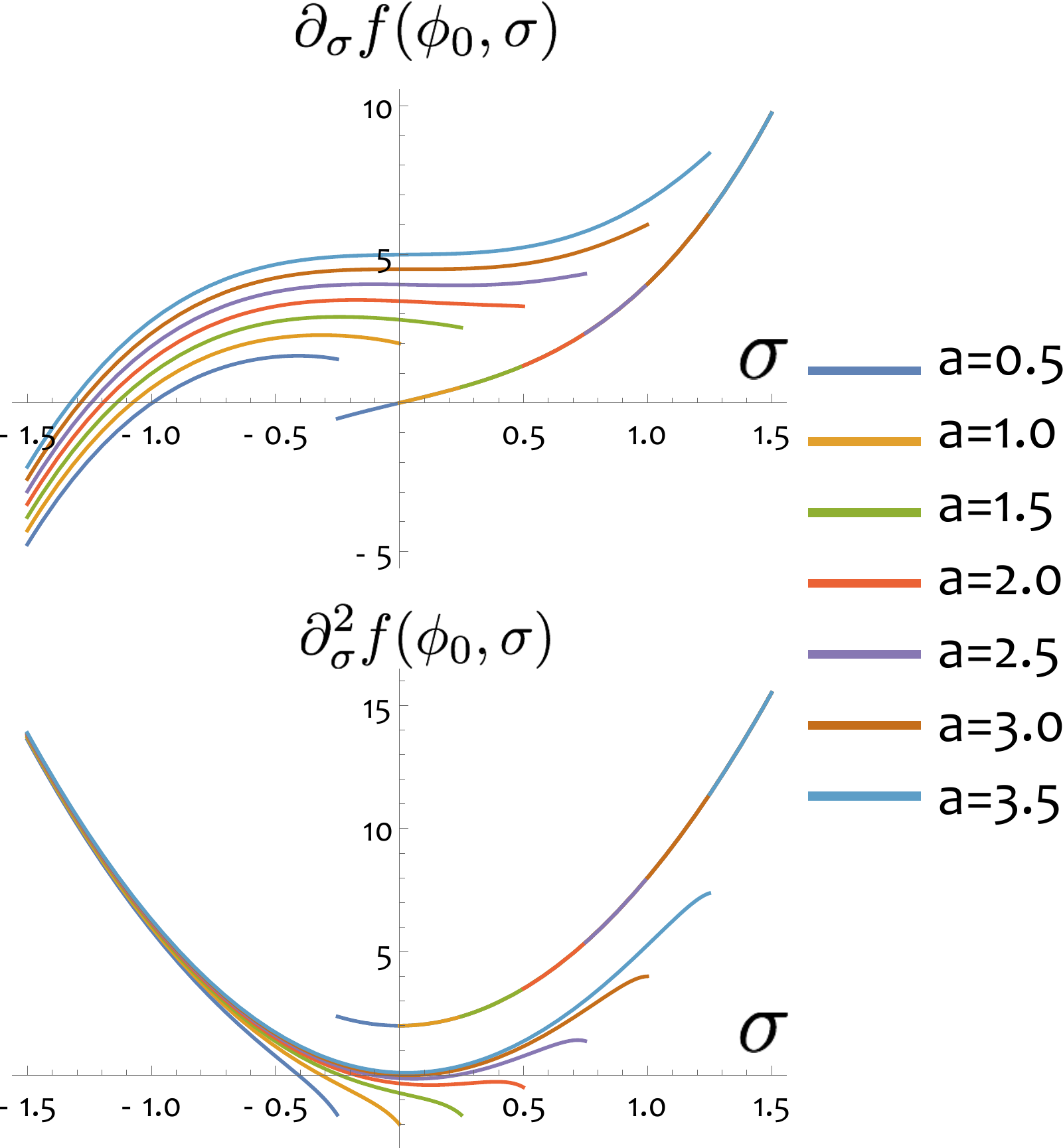}
	\caption{First and second derivative with respect to the conserved order parameter $\sigma$ of free energy $f(\phi_0,\sigma)$ plotted in Fig.~S\ref{fig:fe_f_phi0}D. }
	\label{fig:Dsigma_fe}
\end{figure}

The region where the uniform phase $\sigma(x,t)=\sigma(t)$ is linearly unstable is identified as the ``spinodal'' region and can be readily found by identifying the location of the inflexion points of $f(\phi_0,\sigma)$ (see Fig.~S\ref{fig:Dsigma_fe}). In other words one needs to solve
\begin{equation}\label{eq:spinodalcondition}
\partial_\sigma^2 f(\phi_0,\sigma)=0 \, .
\end{equation}
 
\begin{figure}[h!]
	\centering
	\includegraphics[width=0.45\textwidth]{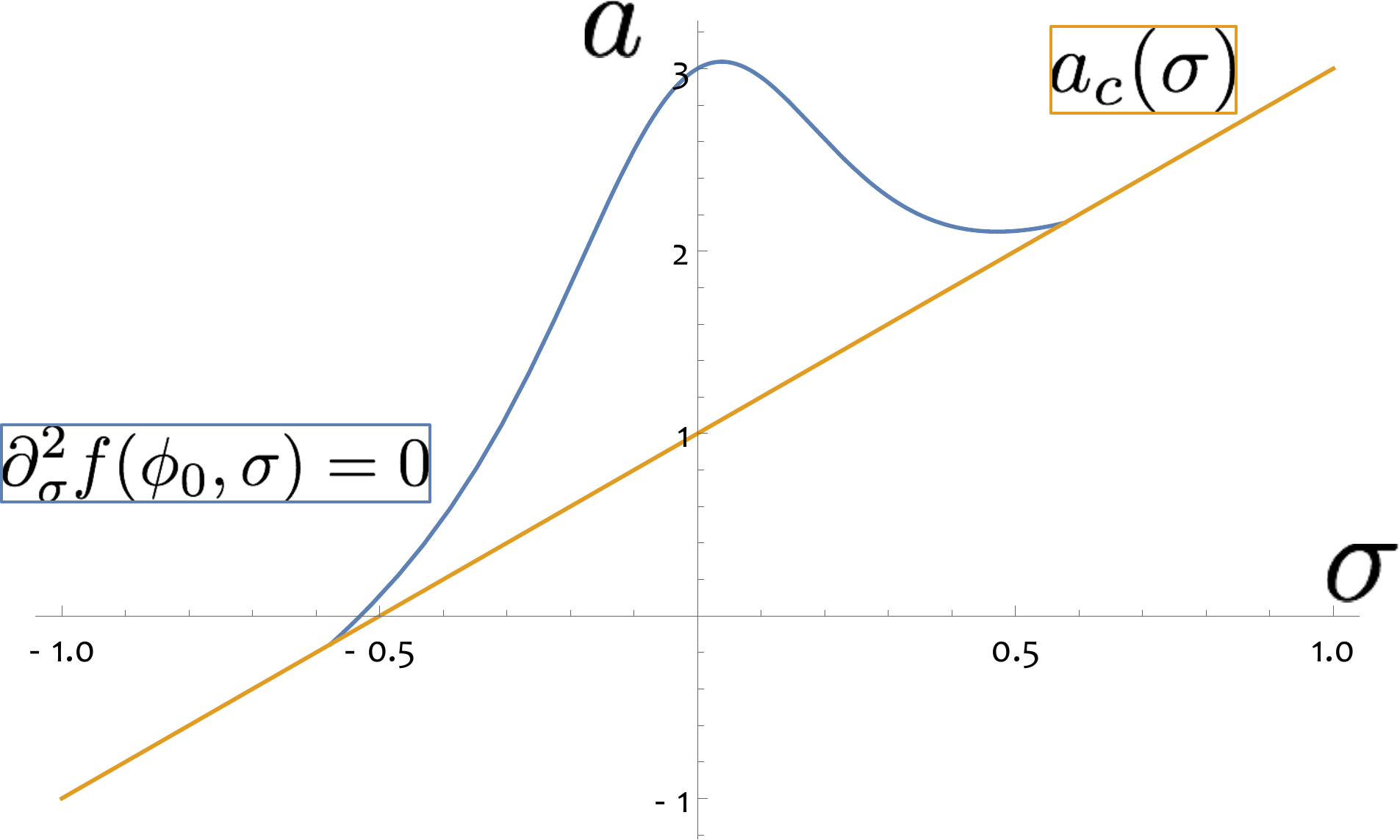}
	\caption{Plot of the spinodal line $a_s(\sigma)$ from eq.~\eqref{eq:spinodal} and of the critical temperature $a_c(\sigma)$ from eq.~\eqref{eq:ac}. }
	\label{fig:spinodal}
\end{figure}

This condition gives an analytical expression for the spinodal lines (for $\chi=2$) as (see Fig.S~\ref{fig:spinodal})
\begin{equation}
a_s(\sigma) = \dfrac{72 \sigma^5 + 27\sigma^4 +24 \sigma^3 + 9 \sigma^2 + 2 \sigma + 3}{36 \sigma^4 + 12 \sigma^2+1} \label{eq:spinodal}\\
\end{equation}
valid if $-0.57735 < \sigma_0 < 0.57735$.

The range of values of $\chi$ for which the requirement in Eq.~\eqref{eq:spinodalcondition} returns a solution $a_s(\sigma)$ is $1.72 \leq \chi \leq 2.67$. For this reason we chose a value in this range, i.e. $\chi=2$.



%

When the system undergoes phase decomposition in the conserved field $\sigma$, two coexisting solutions (or phases) appear, with distinct values of supercoiling: we label $\sigma_+$ the higher supercoiling solution and $\sigma_-$ the lower supercoiling transition (in general $\sigma_+$ and $\sigma_-$ need not have opposite sign). he values of $\sigma_+$ and $\sigma_-$ are found by the common tangent construction, i.e. by equating the chemical potentials 
\begin{equation}
\mu(s) = \left. \dfrac{\partial f(\phi_0,\sigma)}{\partial \sigma}\right|_s
\end{equation}
and the pressures $\Pi(s)$ 
\begin{equation}
\Pi(s) = f(\phi_0,s) - \left. \dfrac{\partial f(\phi_0,\sigma)}{\partial \sigma} \right|_s s \,
\end{equation}
in the two phases, thereby leading to a system of two equations in two unknowns:
\begin{align}
&\mu(\sigma_-)=\mu(\sigma_+)\label{eq:commontangent_mu} \\ 
&\Pi(\sigma_-)=\Pi(\sigma_+) \label{eq:commontangent_pi} \,.
\end{align}
Equation \eqref{eq:commontangent_mu} states that the chemical potentials of the two phases must balance in order for the system to be in equilibrium. Equation \eqref{eq:commontangent_pi} restricts the translational degree of freedom and it identifies the only pair of points of $f(\phi_0,\sigma)$ that have equal tangent $\mu(\sigma_+)=\mu(\sigma_-)$ and that can be joined by a straight line. 

\begin{figure}[t]
	\centering
	\includegraphics[width=0.45\textwidth]{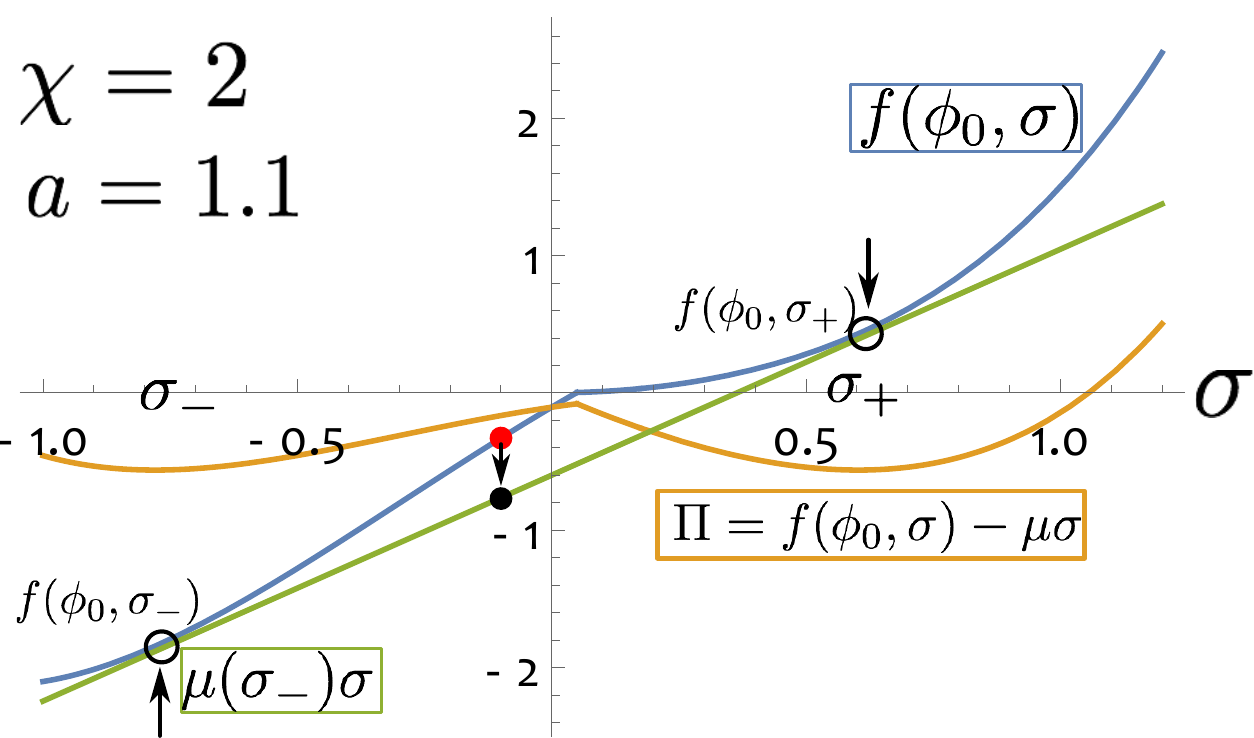}
	\caption{Graphic common tangent construction. The blue curve is the free energy $f(\phi_0,\sigma)$. The green curve is the term $\mu(\sigma_+)\sigma$ where $\mu(\sigma_+)=\mu(\sigma_-)=\left. df/d\sigma\right|_{\sigma_-}$. The orange curve is the pressure $\Pi(\sigma)=f(\phi_0,\sigma)-\mu(\sigma_+)\sigma$. The red dot is the value of the free energy density in the uniform phase $f(\phi_0,\sigma_0=-0.1)$, while the black dot gives the value of the free energy per unit volume in the phase separated state, i.e. $f=f_+ f(\phi_0,\sigma_+)+f_- f(\phi_0,\sigma_-)$ (see text). }
	\label{fig:commtang}
\end{figure}

Graphically, this procedure can be summarised as in Figure~S\ref{fig:commtang}. The blue curve is the free energy $f(\phi_0,\sigma)$ resulting from a choice of $a=1.1$. The green line corresponds to $\mu(\sigma_-)\sigma=\mu(\sigma_+)\sigma$ and the orange curve to the pressure $\Pi(\sigma)$. The common tangent construction identifies the points $\sigma_-$ and $\sigma_+$ with same tangent $\mu$ and with equal pressure, i.e. where the value of the tilted free energy density (orange curve) is the same.  The red dot placed along the (blue) curve $f(\phi_0,\sigma)$ represents the value of the free energy density $f(\phi_0,\sigma_0)$ for $\sigma_0=-0.1$, which is higher than the free energy per unit volume of the phase-separated system:
\begin{equation}
f_+ f(\phi_0,\sigma_+) + f_- f(\phi_0,\sigma_-)
\label{eq:free_en_decomp}
\end{equation}
represented by the black dot. In eq.~\eqref{eq:free_en_decomp}, $f_+$ and $f_-$ are the fractions of the system in the high and low supercoil phases, respectively.

\begin{figure}[b]
	\centering
	\includegraphics[width=0.45\textwidth]{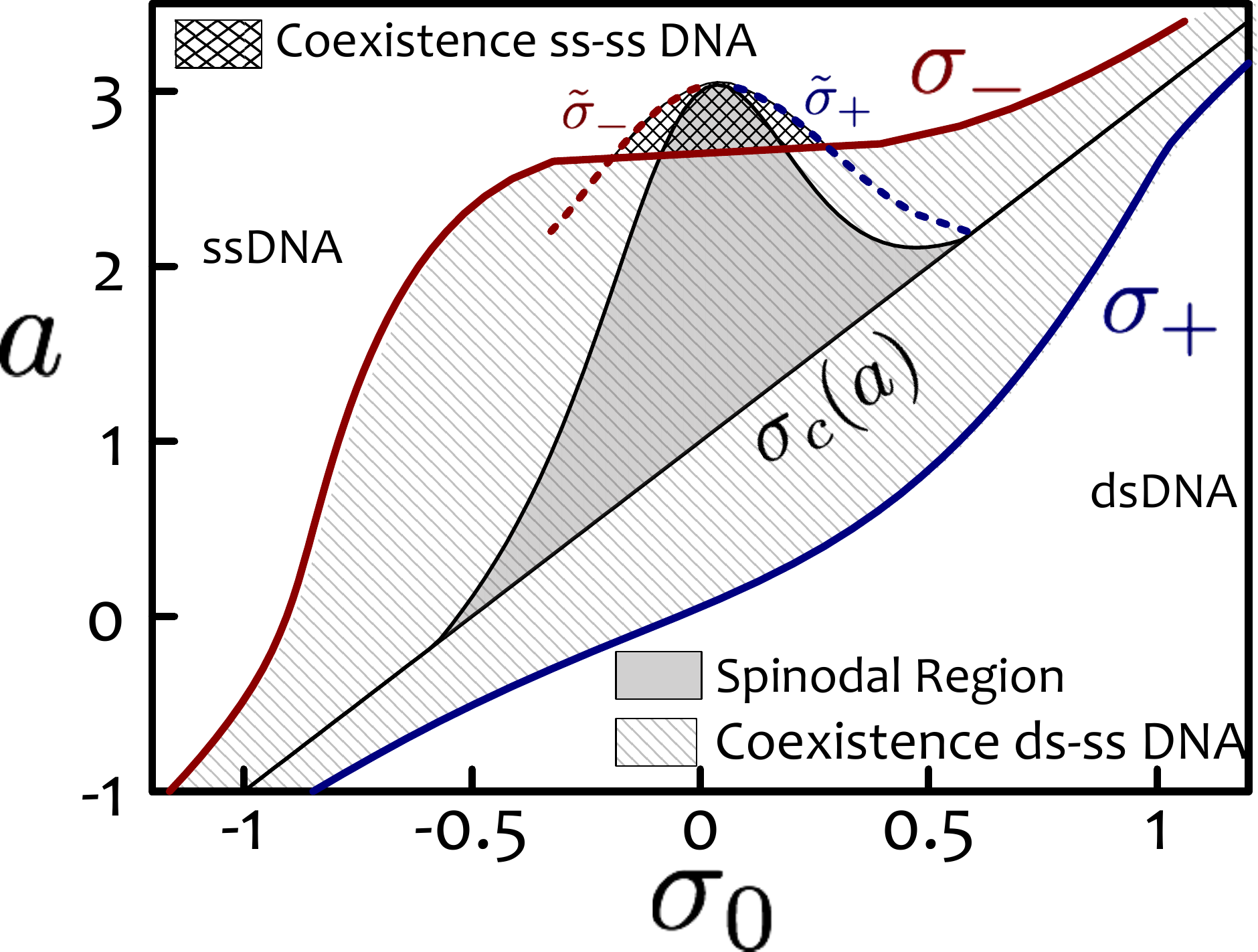}
	\caption{Phase diagram reporting the binodals $\sigma_-(a)$ and $\sigma_+(a)$ obtained by numerically solving eqs.~\eqref{eq:commontangent_mu} and \eqref{eq:commontangent_pi}. The critical line $\sigma_c(a)$ is also shown to be ``hidden'' by the coexistence region.  }
	\label{fig:binodals}
\end{figure}

For a general value of the initial supercoiling field $\sigma_0$, one can find $f_+$ and $f_-$ through the so-called lever-rule:
\begin{align}
&f_+ + f_-= 1 \\ 
&f_- \sigma_- + f_+ \sigma_+=\sigma_0
\end{align}
which gives
\begin{equation}
f_- = \dfrac{ \sigma_+ -\sigma_0}{\sigma_+ - \sigma_-} \,
\end{equation}
and
\begin{equation}
 f_+ = \dfrac{\sigma_0 - \sigma_-}{\sigma_+ - \sigma_-} \, .
\end{equation}

Equations \eqref{eq:commontangent_mu} and \eqref{eq:commontangent_pi} can be solved numerically and give the binodal lines $\sigma_+(a)$ and $\sigma_-(a)$. These are plotted in Fig.~S\ref{fig:binodals} along with the critical line $\sigma_c(a)$ and the spinodal line from eq.~\eqref{eq:spinodal}. 

As one can notice, the binodal line $\sigma_-(a)$ crosses through the spinodal region leaving a reentrance where the uniform solution is linearly unstable, yet the binodals $\sigma_+(a)$ and $\sigma_-(a)$ are not the stable phases. In this region two stable open (ssDNA) phases are observed, whose supercoiling take values on additional binodals $\tilde{\sigma}_+$ and $\tilde{\sigma}_+$ shown in dashed lines.

\section{Equations of Motion: Model C}   

The equations describing the dynamics of the non-conserved field $\phi$ and the conserved one $\sigma$ can be derived from the Landau free energy as~\cite{ChaikinLubensky}
\begin{align}
&\dfrac{\partial \phi(x,t)}{\partial t}= - \Gamma_\phi \dfrac{\delta \mathcal{H}(\phi,\sigma)}{\delta \phi}\notag \\ 
&\dfrac{\partial \sigma(x,t)}{\partial t}= \Gamma_\sigma \nabla^2 \dfrac{\delta \mathcal{H}(\phi,\sigma)}{\delta \sigma} \label{eq:modelC}
\end{align}
where $\Gamma_\phi$ and $\Gamma_\sigma$ are relaxation constant, and $\mathcal{H}$ denotes the total free energy, 
\begin{equation}
	\mathcal{H}(\phi,\sigma)=\int \left( f(\phi,\sigma)+ \gamma_\phi \left(\nabla \phi\right)^2 + \gamma_\sigma \left(\nabla \sigma\right)^2 \right) dx \,.\\
\end{equation}
The equations of motion are therefore:
\begin{align}
&\dfrac{\partial \phi(x,t)}{\partial t}= \notag\\
&=- \Gamma_\phi \left[  2\left( 1 - a + \dfrac{b^2}{2 c}\right)\phi + 3 b \phi^2 + 4 c \phi^3 + 2 \chi \phi \sigma - \gamma_\phi \nabla^2 \phi \right] \notag \\ 
&\dfrac{\partial \sigma(x,t)}{\partial t}= \Gamma_\sigma \nabla^2 \left[ 2 a_\sigma \sigma + 4 b_\sigma \sigma^3 + \chi \phi^2 - \gamma_\sigma \nabla^2 \sigma \right] \, . \label{eq:modelC2}
\end{align}
Equations~\eqref{eq:modelC2} are solved numerically on a 1D lattice of size $L=1000$ via Euler method with integration time $dt=10^{-3}$ $t$ and lattice spacing $dl=1$. The parameters are set as before: $b=-4$, $c=-b/2$, $a_\sigma=1$ and $b_\sigma=1/2$. The surface tensions $\gamma_\sigma$ and $\gamma_\phi$ are both set to $10$ in units of squared lattice spacing. The mobility $\Gamma_\sigma$ is set so that the supercoiling field $\sigma$ relaxes more quickly than $\phi$ (as we expect to be the case for DNA during melting), i.e.  $\Gamma_\sigma=10$ (units $dl^2 t^{-1}$)  while  $\Gamma_\phi=1$ (units $t^{-1}$).

The system is initialised with a small denatured bubble of size $l_b = 10 \ll L$ so that
\begin{equation}
\phi(x,0) = 
\begin{cases}
1 \text{ for } x \in \text{bubble}\\
0 \text{ otherwise} \\
\end{cases}
\end{equation} 
while the supercoiling field is set to
\begin{equation}
\sigma(x,0)=\begin{cases}
 -1 \text{ for } x \in \text{bubble}\\
 \dfrac{L \sigma_0 + l_b}{L-l_b}  \text{ otherwise} \, .
\end{cases}
\end{equation}
Finally, since the molecule is circular-closed, we set periodic boundary conditions along $L$ for the values of the fields and their derivatives, i.e. $\phi(-L/2,t)=\phi(L/2,t)$, $\partial_x \phi(-L/2,t)= \partial_x \phi(L/2,t)$ and $\sigma(-L/2,t)=\sigma(L/2,t)$, $\partial_x \sigma(-L/2,t)= \partial_x \sigma(L/2,t)$ for any time $t$.

Importantly, we can model topologically unconstrained molecules by setting $\chi=0$, for which the denaturation field evolves uncoupled from $\sigma$. This scenario can be viewed as an approximation for nicked molecules where the supercoiling relaxes infinitely fast (through strand rotations at the nick).  

\begin{figure*}[t]
	\centering
	\includegraphics[width=\textwidth]{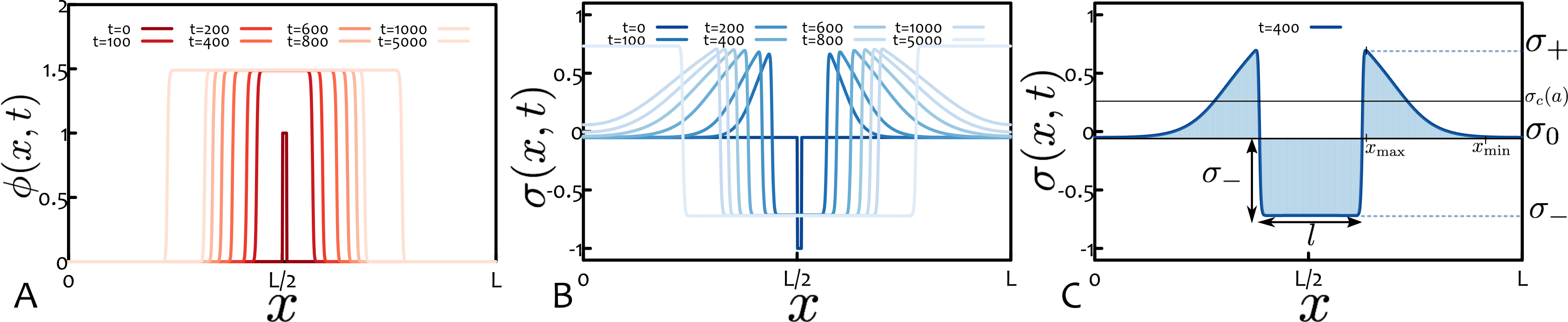}
	\caption{(A-B) Evolution of the fields $\phi(x,t)$ and $\sigma(x,t)$ as a function of position $x$ starting from a small denatured bubble in the middle of the chain. Panel (C) highlights the area within the denatured bubble ($=l(t) \sigma_-$) which must equal the area outside it; in other words, the supercoiling wave forming outside the growing bubble has an area that grows linearly with $l(t)$. Because the maximum of the wave is roughly constant at $\sigma_+$, its base must then grow as $l(t)$. For this reason, one can approximate the slope of the ``front'' as $\left. \partial_x\sigma(x,t)\right|_{\rm front} \simeq (\sigma(x_{\rm max},t)-\sigma(x_{\rm min},t))/(x_{\rm max}(t)-x_{\rm min}(t)) \simeq (\sigma_+ - \sigma_0)/l(t)$. Parameters: $a=1.5$, $\chi=2$ and $\sigma_0=-0.06$.}
	\label{fig:fieldprofile}
\end{figure*}

Starting the simulations from the above-mentioned conditions, we can evolve the system and record the dynamics of the fields in time. The typical profiles are reported in Fig.~\ref{fig:fieldprofile}(A,B), where the denaturation field $\phi$ and supercoiling field $\sigma$ are shown for $a=1.5$, $\chi=2$ and $\sigma_0=-0.06$. For these values of $\sigma_0$ and $a$, the uniform solution is (linearly) unstable and for this reason the initially small denatured bubble grows in time and eventually arrests at the point where two stable phases coexist. At the same time, the supercoiling field is ``emptied'' from inside the bubble, where it takes the stable value of $\sigma_-$, and ``poured'' in the $L-l(t)$ portion of the chain, where it eventually attains the value $\sigma_+$.  

As discussed in the main text, the size of the denatured bubble $l(t)$ has been found to scale in time as 
\begin{align}
&l(t) \sim t^{1/2} \hspace{0.2 cm}\text{ for topologically constrained DNA}\\
&l(t) \sim t^{1} \hspace{0.5 cm} \text{ for topologically unconstrained DNA}\, . 
\end{align}
The exponent for unconstrained DNA, such as nicked or linear, can be simply explained by assuming that for temperatures above a critical $T_c=\epsilon/\Delta S$ the system can lower its (free) energy for each un-paired (denatured) base-pair as 
\begin{equation}
f \simeq (\epsilon - T \Delta S)l.
\end{equation} 
This directly implies that~\cite{ChaikinLubensky}
\begin{equation}
\dfrac{dl}{dt} \simeq \dfrac{1}{\gamma} \dfrac{d f}{d l} \sim const
\end{equation}
and therefore $l(t)\sim t$.

On the other hand, the value of $\alpha$ observed for topologically constrained, such as circular closed, molecules can be understood by quantifying the slowing down of the denaturation field due to the accumulation of a ``wave'' of supercoiling on either side of the growing region. In particular, one can argue that the flux of $\phi$ through a base pair at the interface of a growing bubble is 
\begin{equation}
J_\phi \sim \phi_{\rm ss} \dfrac{dl}{dt},
\end{equation} 
while the flux of $\sigma$ can be obtained by noticing that the ``wave'' roughly assumes a triangular shape with height $h=\sigma_+-\sigma_0$ and base $b \sim l(t)$, so that the area enclosed is proportional to the one expelled from inside the denatured bubble $\sim l(t)$ (see Fig.~S\ref{fig:fieldprofile})). In light of this one can approximate
\begin{equation}
J_\sigma=-\Gamma_\sigma \partial_x \sigma  \simeq \Gamma_\sigma \dfrac{\sigma_+ - \sigma_0 }{l},  
\end{equation}
for, e.g., the flux in the forward direction.
At equilibrium, the two fluxes must balance, i.e. $J_\phi \sim J_\sigma$, and therefore 
\begin{equation}
\dfrac{dl}{dt} \sim \Gamma_\sigma \dfrac{\sigma_+ - \sigma_0}{\phi_{\rm ss} l} 
\end{equation}
or $l(t) \sim t^{1/2}$.

\begin{figure}[b]
	\centering
	\includegraphics[width=0.45\textwidth]{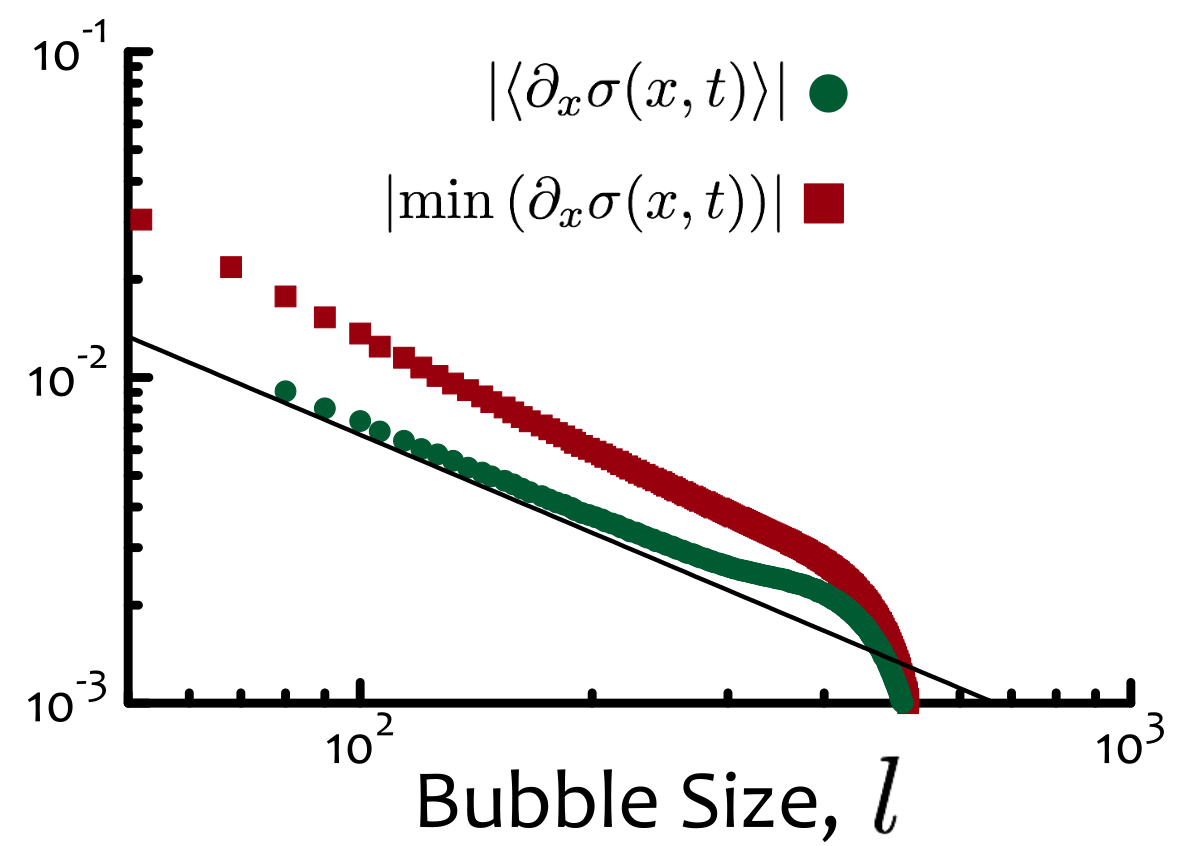}
	\caption{Slope of the $\sigma$ field along the front of the wave forming on either side of a growing bubble. Parameters as in Fig.~\ref{fig:fieldprofile}.}
	\label{fig:slope_sigma}
\end{figure}

Because the approximation $\left. \partial_x \sigma(x,t)\right|_{\rm front}  \simeq (\sigma_+ - \sigma_0)/l(t)$ might look crude, we numerically computed the minimum and the average slope attained by the $\sigma$ field on the front of the supercoiling wave. These two quantities are plotted in Fig.~S\ref{fig:slope_sigma} against the size of the bubble $l$. From the figure one can readily notice that the approximation is in fact not far from the real behaviour. 

\section{Computing Linking number in denatured bubbles}
For a topologically constrained dsDNA the linking number $Lk$ has to be preserved at any time. In particular, even when a circular closed molecule of DNA is fully denatured, all the initial linking number has still to be preserved. One may imagine this situation as an extreme case where the only denaturation bubble takes up the whole DNA molecule. In light of this, one may ask whether a generic bubble of size $l$ can store some non-zero linking number, $Lk_d$. This case is generally discarded in simple models~\cite{Kabakcioglu2009a,Jeon2010,Poland1966a}, which consider no supercoiling within denatured regions. On the contrary, arguments from elasticity theory~\cite{Benham1979} suggest and support the idea that some linking may still be present within denatured bubbles, since a complete expulsion of this would then contribute to increase the torsional (twist) or bending (writhe) energies of the intact (non-denatured) segments through the well-known $Lk=Wr+Tw$ relation~\cite{Fuller1978}.   

\begin{figure}[t]
	\centering
	\includegraphics[width=0.4\textwidth]{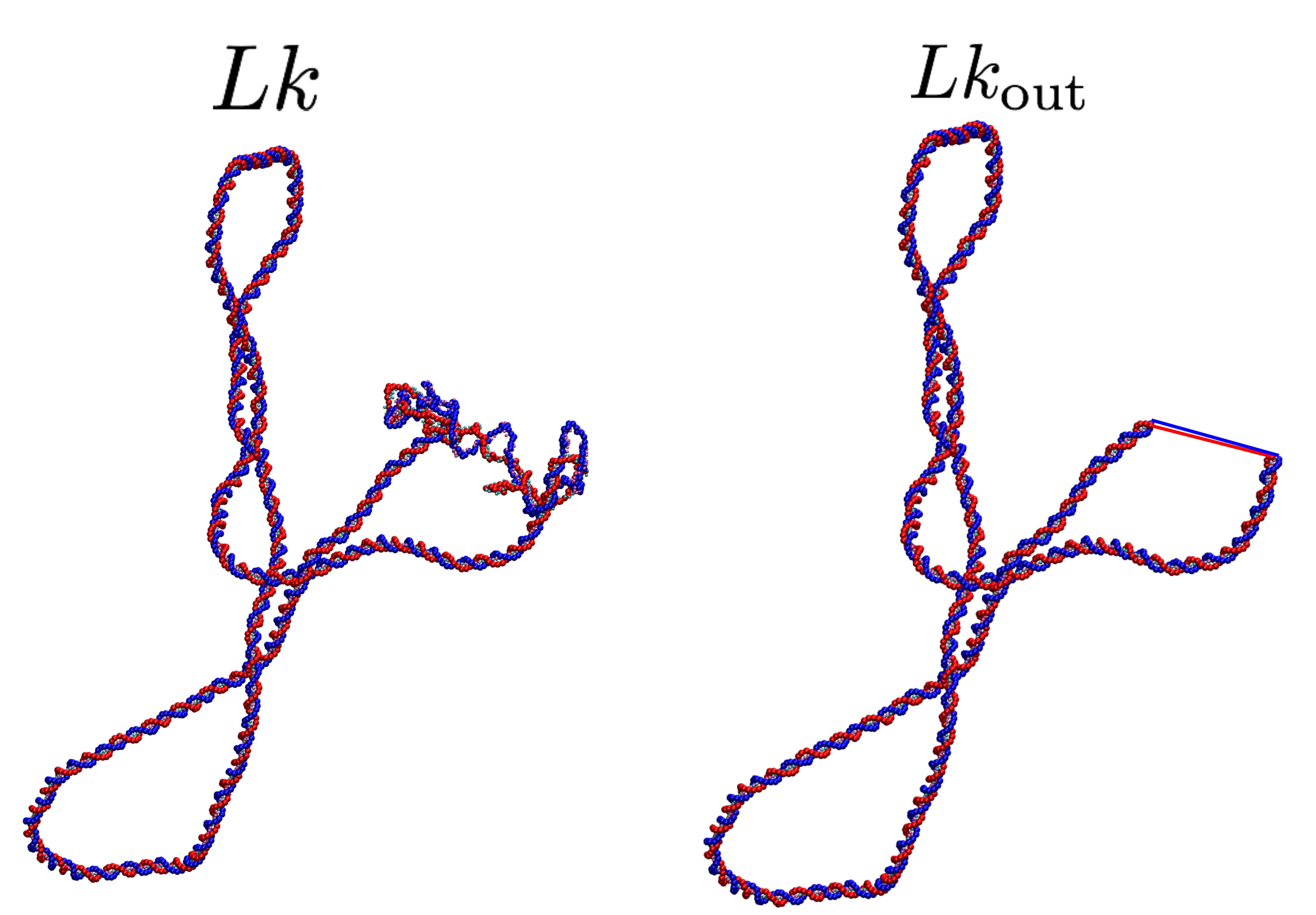}
	\caption{\textbf{Computing linking number in denatured molecules} Sketch of the process through which we compute $Lk_{d}$. We take a denatured molecule with fixed bubble size $l$ (here $l=200$ bp) and replace the denatured strands with straight lines joining the first non-denatured beads on either side of the bubble. Then, we compute the linking number outside the bubble ($Lk_{\rm out}$) using Eq.~\eqref{computelk} and derive the linking stored within the bubble as $Lk_d=Lk-Lk_{\rm out}$.}
	\label{sketchbubble}
\end{figure}

In order to test this argument we performed simulation in which we kept the strength of the hydrogen bond $\epsilon_{HB}=6.0 k_BT$ constant and high enough so denaturation could not take place. Then, for rings with different level of supercoiling ($\sigma = 0.06, 0, -0.06$, or $Lk=106,100,94$, respectively), we created a bubble of size $l$ by manually deleting the corresponding consecutive hydrogen-bonds along the chain. 

After equilibration we computed the linking number by numerically calculating 
\begin{equation}
Lk(C_{R},C_{B}) = \frac{1}{4\pi} \displaystyle \int_{C_{R}} \int_{C_{B}} \frac{\bm{r}_{R} - \bm{r}_{B}}{\vert \bm{r}_{R} - \bm{r}_{B}\vert^{3}} \cdot (d\bm{r}_{R} \times d\bm{r}_{B}),
\label{computelk}
\end{equation}
between curves $C_{R}$ and $C_{B}$. In Eq.~\eqref{computelk}, $\bm{r}_{R}$ and $\bm{r}_{B}$ are the vectors defining the position of the beads along the curves $C_{R}$ and $C_{B}$, respectively. 

\begin{figure}[t]
	\vspace{0.4 cm}
	\centering
	\includegraphics[width=0.4\textwidth]{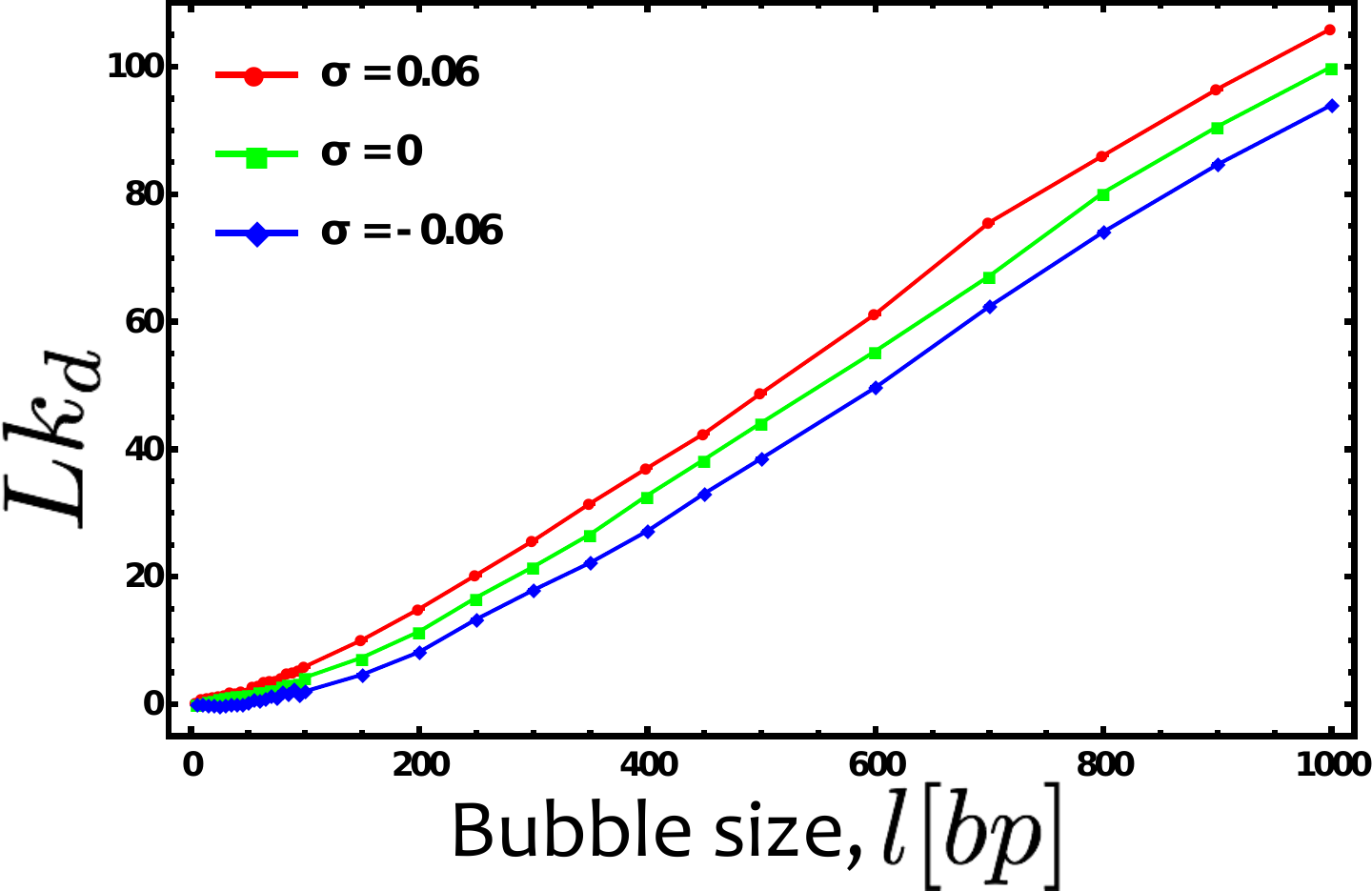}
	\caption{\textbf{$Lk_d$ grows with the size of the denatured bubble.} The figure shows that the linking number stored within the denatured region, $Lk_d$, grows as function of its length $l$. The curves (red, green and blue) correspond to circular closed molecules with different levels of supercoiling: $\sigma = 0.06, 0, -0.06$, or $Lk=106,100,94$, respectively.}
	\label{Lkbubble_vs_length}
\end{figure}

In order to find the linking number \emph{inside} the denatured region, $Lk_d$, we first replaced the bubble by straight lines joining the first non-denatured base-pairs (see Fig.~S\ref{sketchbubble}), so that Eq.~\eqref{computelk} returned the linking number \emph{outside} the denatured region, $Lk_{\rm out}$. Finally, we simply use the invariance of the total linking number to find $Lk_d=Lk - Lk_{\rm out}$ (see Fig.~S\ref{Lkbubble_vs_length}).

\end{document}